\documentclass[12pt]{article}

\usepackage{url}
\usepackage{amsmath,amssymb,amsthm,amscd}
\usepackage{graphicx}

\usepackage{natbib}
\usepackage{algorithm}	
\usepackage{algorithmic}
\usepackage{bm}
\usepackage{bbm}
\usepackage{enumitem}

\usepackage{wrapfig}
\usepackage{lscape}
\usepackage{rotating}
\usepackage{epstopdf}

\usepackage{eurosym}
\usepackage{textcomp}
\usepackage{xcolor}

\usepackage{cprotect}

\usepackage[toc,page]{appendix}

\graphicspath{ {Images/} }

\usepackage{xr} 

\makeatletter

\theoremstyle{plain}

\newcommand{\ts}{^{\sf T}} 

\newcommand{\blind}{1}

\begin{document}

\def\spacingset#1{\renewcommand{\baselinestretch}%
{#1}\small\normalsize} \spacingset{1}


\if1\blind
{
  \title{\bf Additive stacking for disaggregate electricity demand forecasting}
 \author{
    Christian Capezza, Biagio Palumbo, \\
    Department of Industrial Engineering, \\ 
    \vspace{0.4cm}
    Universit{\`a} degli Studi di Napoli Federico II \\
    Yannig Goude, \\
    \vspace{0.4cm}
    {\'E}lectricit{\'e} de France R\&D \\
    Simon N. Wood and Matteo Fasiolo \\
    School of Mathematics, University of Bristol}
    \date{}
  \maketitle
} \fi

\if0\blind
{
  \bigskip
  \bigskip
  \bigskip
  \begin{center}
    {\LARGE\bf Additive stacking for disaggregate electricity demand forecasting}
\end{center}
  \medskip
} \fi

\bigskip
\begin{abstract}
Future grid management systems will coordinate distributed production and storage resources to manage, in a cost effective fashion, the increased load and variability brought by the electrification of transportation and by a higher share of weather dependent production. Electricity demand forecasts at a low level of aggregation will be key inputs for such systems. We focus on forecasting demand at the individual household level, which is more challenging than forecasting aggregate demand, due to the lower signal-to-noise ratio and to the heterogeneity of consumption patterns across households. We propose a new ensemble method for probabilistic forecasting, which borrows strength across the households while accommodating their individual idiosyncrasies. In particular, we develop a set of models or `experts' which capture different demand dynamics and we fit each of them to the data from each household. Then we construct an aggregation of experts where the ensemble weights are estimated on the whole data set, the main innovation being that we let the weights vary with the covariates by adopting an additive model structure. In particular, the proposed aggregation method is an extension of regression stacking \citep{breiman1996stacked} where the mixture weights are modelled using linear combinations of parametric, smooth or random effects. The methods for building and fitting additive stacking models are implemented by the \verb|gamFactory| R package, available at \url{https://github.com/mfasiolo/gamFactory}.
\end{abstract}

\noindent%

\spacingset{1} 



\section{Introduction} \label{sec:introduction}

The electricity grid is transitioning from a system with centralised production and limited storage, both controlled by the system operator or other industrial entities, to a more complex setting where production and storage are decentralised, and the former is strongly weather dependent. The transition is motivated by the need to reduce carbon emissions, which is leading to a shift from fossil fuel to renewable power production and to the electrification of the transportation system.
These developments represent a challenge for current grid management systems, as it will be necessary to satisfy the extra demand generated by a large fleet of electric vehicles in a context where production is less flexible and more uncertain. 
To limit the need for expensive infrastructural works, aimed at increasing the physical capacity of the electricity network, intelligent grid management systems and policies must be put in place. For example, dynamic electricity pricing and remotely controlled consumption can be used as demand-side tools to reduce the daily demand peak and to coordinate demand with time-varying renewable energy production. 

Electricity demand forecasts at the system-wide or regional scale are key inputs for production planning and grid management under the current, centralised, electricity system. As the availability of distributed production and storage increases, demand forecasts at a lower level of aggregation will become more important. To illustrate this, we consider a simple scenario, outlined here, and described in detail in Section \ref{sec:peakShave}. Consider a portfolio of residential customers, whose demand is recorded half-hourly via smart meters. Each household is equipped with a home battery and its charge/discharge schedule is determined, one day ahead, by minimising the expected cost to the customer. We assume that the customers are charged based on a composite tariff, with a baseline price for the total amount of energy used and a much higher price for the daily maximum demand. We adopt a daily-max tariff because \cite{pimm2018time} demonstrate that standard time-of-use tariff could lead to little or no reduction in the daily peak demand, as many batteries might start charging simultaneously at the start of the overnight off-peak price. They speculate that a daily-max tariff might be more effective for peak demand shaving, which is a key goal as the network infrastructure must be able to satisfy peak demand and the cost of network reinforcement is expected to reach up to £36bn by 2050 in the UK \citep{pudjianto2013smart}. 

\begin{figure} 
\centering
\includegraphics[scale=0.43]{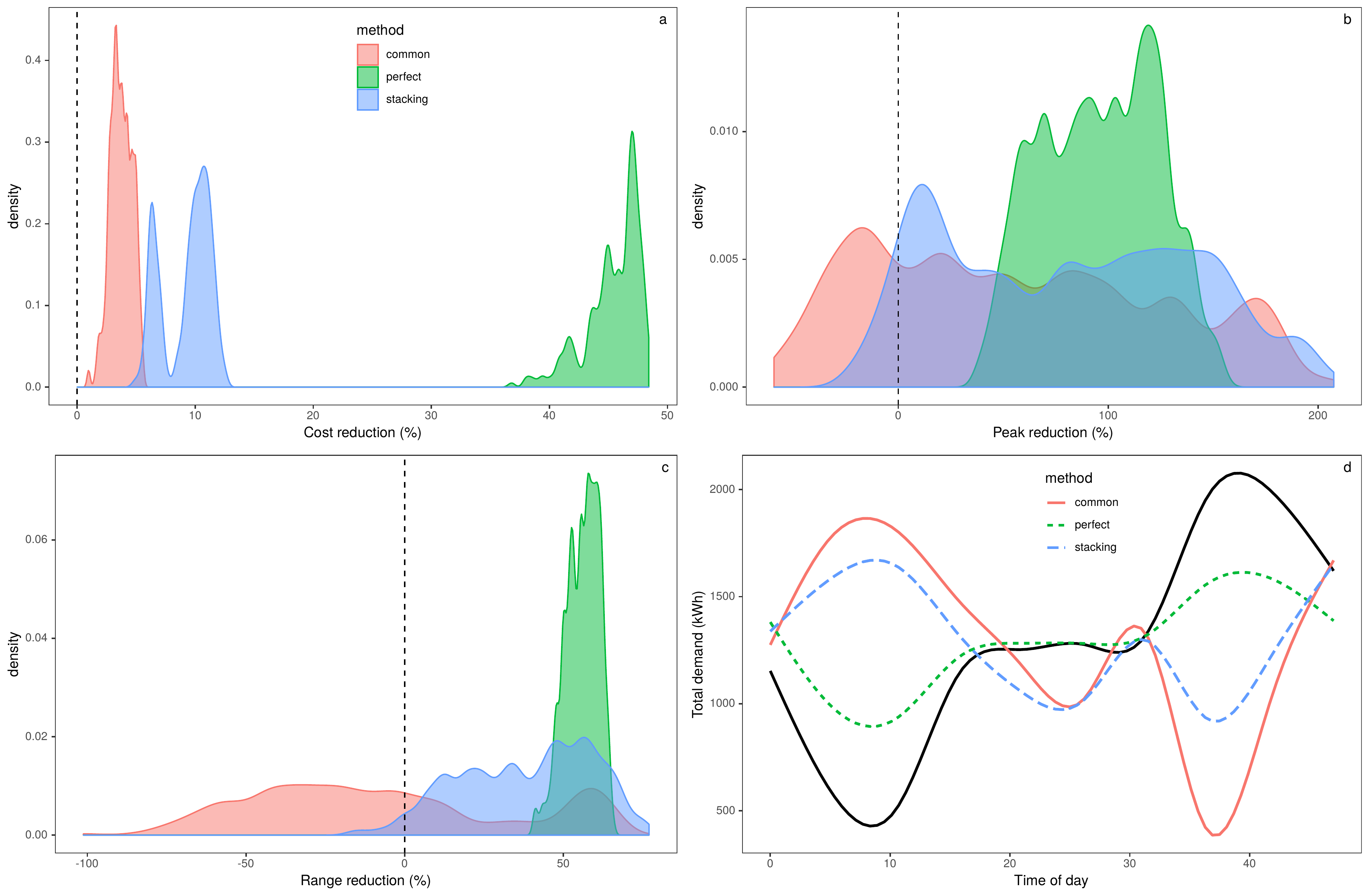} 
\cprotect\caption{Plots a-c show the reduction in the daily total cost of electricity, maximum and range of aggregate demand, relative to leaving the batteries idle, obtained by using different household specific demand forecasts as inputs to the battery optimisation algorithm. The black solid curve in plot d is the average profile of the total demand during weekdays and the other three curves show how the profile is transformed by the batteries charging/discharging schedules corresponding to each forecast. The plots correspond to scenario $A = 0$\euro, described in Section \ref{sec:peakShave}.}
\label{fig:peak_plot}
\end{figure}

\begin{figure} 
\centering
\includegraphics[scale=0.48]{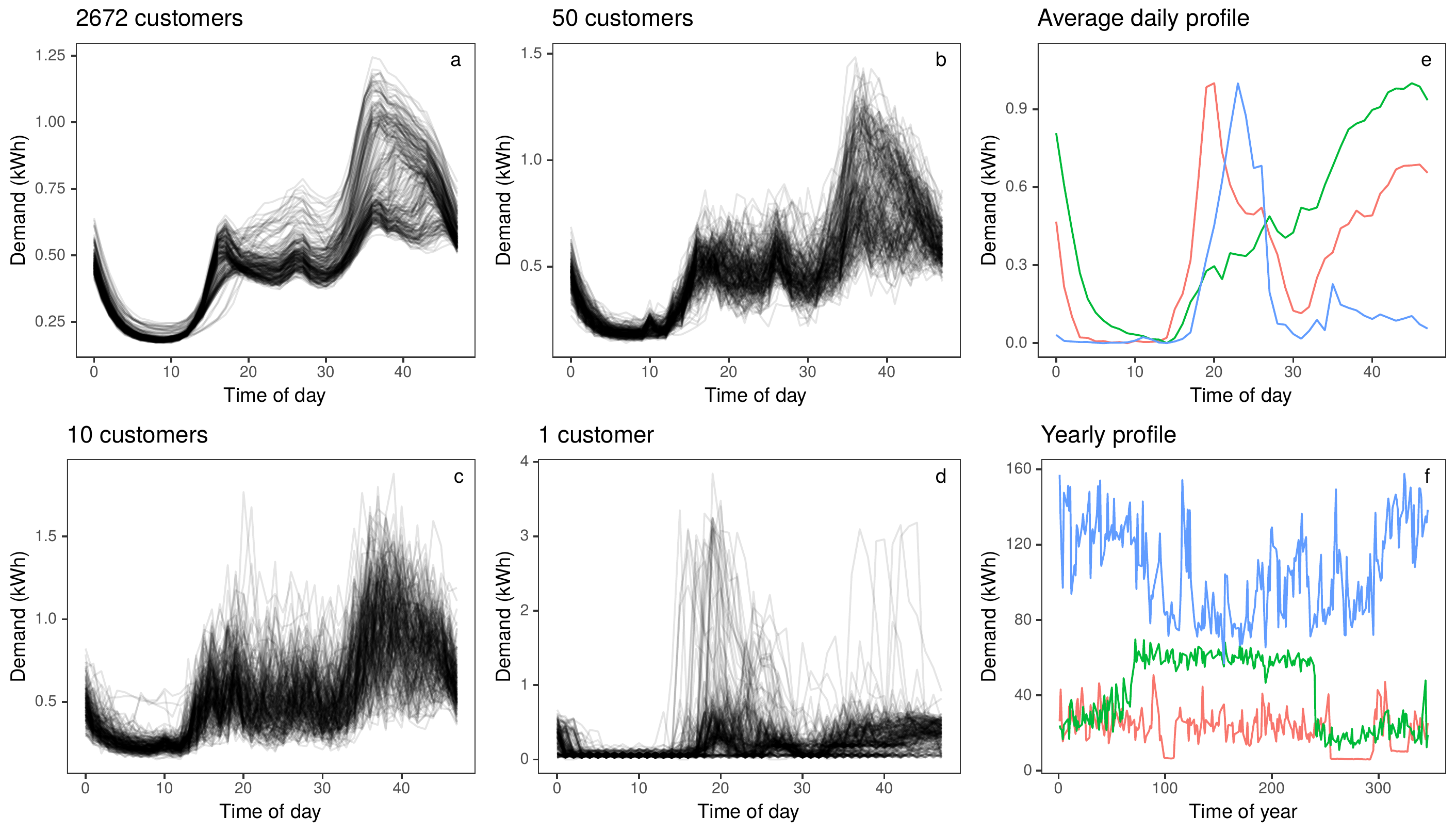} 
\cprotect\caption{Plots a-d show the daily profiles of the demand averaged over increasingly small groups of customers from the CER trial \citep[][see Section \ref{sec:demandForecast} for details]{commission2012cer}.
Plots e and f show the average daily and yearly demand profiles of three customers. The blue profile in plot f has been vertically shifted for visibility.}
\label{fig:Irish_prof}
\end{figure}

Figure \ref{fig:peak_plot} shows that, in the setting just outlined, using accurate household specific forecasts for battery optimisation can lead to substantial cost reductions for the customers and to a flatter aggregate demand profile. In particular, Figure \ref{fig:peak_plot}a shows that households energy costs could be reduced by almost 50\%, relative to no battery usage, under a `perfect' forecast which assumes that households demand is known one day in advance. The plots in Figure \ref{fig:peak_plot}b and \ref{fig:peak_plot}c show that the daily peak and range, the latter being the difference between the maximum and minimum demand, can also be reduced substantially, leading to the flatter aggregate demand profile shown in Figure \ref{fig:peak_plot}d. However, the cost savings and the demand profile flattening just described rely on the use of perfect forecasts at the household level to optimise battery schedules, and Figure \ref{fig:Irish_prof} shows that predictive accuracy is destined to deteriorate with the level of granularity. In particular, plots \ref{fig:Irish_prof}a to \ref{fig:Irish_prof}d show that, while the daily profile is smooth when demand is averaged across the customers, disaggregating the demand leads to rough, less predictable profiles. The low signal-to-noise ratio characterising individual household demand suggests a modelling strategy based on predicting the data from several customers using a single model, to reduce the noise. However, plots \ref{fig:Irish_prof}e and \ref{fig:Irish_prof}f show that the behaviour of customers is highly heterogeneous, hence na{\"i}ve aggregation would induce high bias. To demonstrate this, Figure \ref{fig:peak_plot} shows that the result of using a `common' forecast, which simply scales a common predicted daily demand profile depending on the household characteristics (see model $M_4$ in Section \ref{sec:dat_and_exp} for details), are poor. In particular, the battery schedules derived under such a forecast lead to little cost savings, because minimising the cost to the customer under a daily-max tariff requires predicting the profile of each household. Furthermore, rescaling the same common profile to forecast the demand of each customer leads to battery schedules that are highly correlated, hence there is no reduction in the daily range of the aggregate demand (see Figure \ref{fig:peak_plot}c). 

To take into account the heterogeneity of demand dynamics across households, we fit a set of statistical models or `experts' separately to each household. The experts are designed to capture different aspects of individual household demand, such as the smooth daily demand profiles of Figure \ref{fig:Irish_prof}e and the abrupt change-points of Figure \ref{fig:Irish_prof}f. To alleviate the fact that household demand data is characterised by a low signal-to-noise ratio, we `borrow information' across households by constructing a weighted combination of experts, where the weights are estimated by using the data from all the households to fit a single aggregation model. The key methodological innovation is that the weights of the experts can depend semi-parametrically on covariates such as the day of the week, the time of day, household characteristics and so on. The effect of the covariates on the weights is modelled additively, that is using linear combinations of parametric and smooth effects based on spline basis expansions. The results in Section \ref{sec:demandForecast} show that the forecast produced by the aggregation model is more accurate than those obtained under any of the experts and that it leads to better battery scheduling in the example application.

We call the proposed aggregation method `additive stacking', because it is an additive extension of regression stacking \citep{breiman1996stacked}. Closely related work is that of \cite{yao2018using}, who use stacking to average Bayesian predictive distributions, thus producing full probabilistic forecasts. Similarly to them, we model the full distribution of household demand, rather than just its mean. But, given that our stacking model is more complex and that the data set we consider is quite large, we do not adopt a full Bayesian framework based on Markov chain Monte Carlo (MCMC) sampling as done by \cite{yao2018using}, but we fit the model using the approximate empirical Bayes methods of \cite{wood2016smoothing}, which rely on direct optimisation methods. In particular, the regression coefficients are estimated using maximum a posteriori (MAP) methods while the smoothing parameters, which control the wiggliness of the smooth effects, are selected by maximising a Laplace approximation to the marginal likelihood (LAML). We are able to adopt the likelihood based fitting methods of \cite{wood2016smoothing}, aimed at generalised additive models \citep[GAMs,][]{hastie1990generalized}, because we perform stacking in a probabilistic, rather than loss based, context. The parametrisation of the proposed stacking model is non-linear in the regression coefficients, to force the experts' weights to be positive and sum to one. This poses some difficulties when interpreting the estimated effects of the covariates on the weight of each expert. Effective visualisation is essential for GAM model building and checking \citep[see, e.g.,][]{fasiolo2018scalable}, hence we address this difficulty by using the accumulated local effects (ALE) of \cite{apley2016visualizing} to visualise to main effect of the covariates on the aggregation weights. We also show how the uncertainty of the local effect can be quantified at no extra computational cost, by propagating the uncertainty of a Gaussian approximation to the posterior distribution of the regression coefficients.

The proposed additive stacking method is related to several other ensemble methods with varying experts' weights. In particular, many ensemble methods let the weights depend on time. For example, in a time series context, \cite{mcalinn2019dynamic} use a dynamic Bayesian predictive synthesis framework to combine probabilistic forecasts. Similarly, in online mixtures of experts the weights are updated sequentially, as new data becomes available. See \cite{cesa2006prediction} for an overview on online learning and \cite{devaine2013forecasting} for an application to aggregate demand forecasting. The proposed approach differs from such methods because the experts' weights do not depend only on time, but are more flexible semi-parametric functions of all the covariates. The feature weighted linear stacking method of \cite{sill2009feature} is closer to the present proposal, but the weights are modelled using linear combinations of meta-features which must be chosen manually. \cite{coscrato2020nn} proposes a non-linear extension of feature weighted linear stacking, where experts' weights are modelled using neural networks. Hence, the weights depend non-parametrically on the features as in our proposal, but their method focuses on providing point estimates by minimising the quadratic loss, not on modelling the full distribution of the response. Further, by adopting a full probabilistic framework, we are able to fit the model efficiently using the likelihood based framework of \cite{wood2016smoothing} and to provide uncertainty estimates on the fitted model.

The rest of the paper is structured as follows. In Section \ref{sec:method} we present the structure of additive stacking models, we show how they can be fitted efficiently using the direct methods of \cite{wood2016smoothing} and we discuss the use of accumulated local effects to quantify the effect of the covariates on the stacking weight. In Section \ref{sec:demandForecast} we present a set of probabilistic experts which are combined within an additive stacking model, aimed at predicting the individual household demand from the \cite{commission2012cer} trial data. After defining the experts and the stacking model, we demonstrate that the stacked ensemble beats all the individual experts on several loss functions and on the optimal battery scheduling application. Section \ref{sec:conclusion} concludes by summarising the results.

\section{Additive stacking}
\label{sec:method}

\subsection{Model structure}
\label{sec:add_stacking}

According to \cite{leblanc1996combining}, the idea of combining several point estimators to produce a meta-model with improved predictive accuracy dates back at least to \cite{stone1974cross}. The same idea was proposed again by \cite{wolpert1992stacked} under the name of `stacked generalisation', which was then framed and analysed as a linear regression problem by \cite{breiman1996stacked}. To introduce a basic regression stacking setting, let ${\bm y} = \{y_1, \dots, y_N\}$ be a vector of dependent variables, let ${\bm x}_1, \dots {\bm x}_N$ be the corresponding $d$-dimensional vectors of covariates and indicate with $\mathcal{D}$ the data set formed by their pairs, $\{y_i, {\bm x}_i\}$ for $i = 1, \dots, N$. Assume that we have $K$ estimators or `experts', such that $\eta_k({\bm x}_i)$ estimates $\mathbb{E}(y_i|{\bm x}_i)$ under the $k$-th expert. In regression stacking, the experts are combined using a weighted sum, $\sum_{k=1}^K \alpha_k \eta_k({\bm x}_i)$, where the weights are estimated as follows
\begin{equation} \label{eq:basic_stacking}
\hat{\bm \alpha} = \{\hat{\alpha}_1, \dots, \hat{\alpha}_K\} = \underset{\bm \alpha}{\text{argmin}} \sum_{i=1}^N\left\{y_i - \sum_{k=1}^K {\alpha}_k \eta_k({\bm x}_i)\right\}^2, 
\end{equation}
under the constraints $\alpha_k > 0$ and $\sum_{k=1}^K \alpha_k = 1$. According to \cite{breiman1996stacked} the sum-to-one constraint can be omitted, but ignoring the positive weight might hinder predictive accuracy if the experts are strongly correlated. Of course, estimating the experts using $\mathcal{D}$ and then using the same data again to estimate the weights would lead to overfitting. Hence, a cross-validation scheme is typically adopted and $\eta_k({\bm x}_i)$ is obtained by fitting the expert to a subset of $\mathcal{D}$ which excludes the $i$-th data pair. Leave-one-out cross-validation is a simple, but potentially expensive, option for doing this.

Early work on regression stacking focused on combining estimators under the  quadratic loss, but alternative loss functions can be considered. For example, substituting the quadratic loss with the absolute loss leads to a stacked estimator of the conditional median, rather than the mean. The present work is motivated by an electricity demand forecasting application where the full distribution of the response is of interest, hence we consider stacking predictive distributions, not point estimates. In particular, let $p_k(y_i|{\bm x}_i)$ be the $i$-th conditional density estimate produced by the $k$-th expert. Then, probabilistic stacking is performed by forming a mixture, $\sum_{k=1}^K {\alpha}_k p_k(y_i|{\bm x}_i)$, where the weights are estimated by maximising the corresponding log-likelihood, that is
\begin{equation} \label{eq:prob_stacking}
\hat{\bm \alpha} = \underset{\bm \alpha}{\text{argmax}} \sum_{i=1}^N \log \sum_{k=1}^K \alpha_k p_k(y_i|{\bm x}_i), 
\end{equation}
under the constraints mentioned above. As for loss based stacking, a cross-validation scheme must be adopted to avoid overfitting. While \cite{yao2018using} perform additive stacking in a context where the $p_k$'s are Bayesian posterior predictive densities, the stacking methods proposed here are agnostic to the nature of the experts densities, which might be obtained using Bayesian, frequentist or other methods.



In this work, we extend probabilistic stacking by letting the weights vary with the covariates via an additive model structure. In particular, the weights are parametrised as in multinomial logistic regression, that is
\begin{equation} \label{eq:multinomParam}
\alpha_{ki} = \frac{\exp \eta_{ki}}{\sum_{a=1}^{K} \exp \eta_{ai}}, \quad \text{for} \quad k = 2, \dots, K,
\end{equation}
where $\eta_{ki}$ is the linear predictor of the $k$-th expert, evaluated at the $i$-th observation. While $\eta_1$ is fixed to zero for identifiability, the remaining linear predictors are modelled as follows
$$
\eta_{ki} = \sum_{j \in I_k} f_{kj}(\bm x_i), \quad \text{for} \quad k = 2, \dots, K,
$$
where the $f_{kj}$ are parametric, random or smooth effects, based on spline basis expansions, and $I_k$ is the set of indices specifying the effects on which $\eta_{k}$ depends. The smooth effects are constructed using spline basis expansions. In particular, if we drop the indices $j$ and $k$ for notational convenience, we have
$$
f(\bm x_i) = \sum_{l = 1}^{L} b_l(\bm x_i) \beta_{l}, 
$$
where the $b_l$'s are known basis functions and the $\beta_l$'s are unknown regression coefficients, which must be estimated. While the number of basis functions, $L$, is typically chosen to be large enough to avoid over-smoothing, the wiggliness of the effects is controlled by an improper multivariate Gaussian prior on the vector of regression coefficients, $\bm \beta$. The prior is centered at the origin and its precision matrix is ${\bf S}^{\bm \lambda} = \sum_{g=1}^G \lambda_g {\bf S}_g$, where the ${\bf S}_g$'s are positive semi-definite matrices and $\bm \lambda = \{\lambda_1, \dots, \lambda_G\}$ is a vector of positive smoothing parameters. The Bayesian posterior log-density corresponding to such a prior is
\begin{equation} \label{eq:logPosterior}
\log p(\bm \beta|{\bm y}, \bm \lambda) = \sum_{i=1}^N \log \sum_{k=1}^K \alpha_{ki}(\bm \beta) p_k(y_i|{\bm x}_i) - \frac{1}{2} \sum_{g=1}^G \lambda_g \bm \beta \ts {\bf S}_g \bm \beta,
\end{equation}
up an additive constant. Hence the prior log-density is equivalent to a generalised ridge penalty, and increasing the $\lambda_g$'s leads to a posterior which is more concentrated on the null space of the penalty. The null space is spanned by `completely smooth' functions, where the definition of `smooth' depends on the type of prior precision matrix or penalty used. In general, there is no one-to-one correspondence between the effects and the smoothing. For example, the wiggliness of an effect can be controlled via multiple smoothing parameters. 

The additive stacking framework just outlined allows for considerable modelling flexibility, as the whole array of effect types available under standard GAM models can be employed. See \cite{wood2006generalized} for an introduction to splines bases and penalties, in a GAM modelling context. Further, while we indicate with $\bm x$ all the available covariates, it is possible to use different sets of covariates within the experts and to model the stacking weights. The stacking model described in Section \ref{sec:stacking_model} exploits this feature.  
Setting additive stacking in a probabilistic Bayesian framework allows us to employ statistically well-founded and computationally efficient methods for model fitting and inference. In particular, Section \ref{sec:model_fitting} explains how we exploit the methods of \cite{wood2016smoothing} to obtain maximum a posterior (MAP) estimates of the regression coefficients and to select the smoothing parameters using approximate marginal likelihood methods. The ALE visualisation methods described in Section \ref{sec:ale_effects} also benefit from the adoption of a probabilistic Bayesian framework, as the uncertainty of the effects can be quantified using standard asymptotic approximations.

\subsection{Model fitting} \label{sec:model_fitting}

For fixed smoothing parameters, $\bm \lambda$, we obtain MAP estimates of the regression coefficients by maximising the log-posterior (\ref{eq:logPosterior}), using Newton's algorithm. The latter requires the gradient and Hessian of the log-posterior w.r.t. $\bm \beta$, which are provided in Supplementary Material S1 (henceforth SM \ref{app:derivatives}). As for standard GAMs, the real challenge is selecting the smoothing parameters themselves.
We do it by maximising an approximation to the log marginal likelihood, $\mathcal{V}(\bm \lambda) = \log p(\bm \lambda) =  \log \int p(\bm y | \bm \beta)p(\bm \beta|\bm \lambda) d \bm \beta$. In particular, we consider a Laplace approximate marginal likelihood (LAML) criterion
\begin{equation} \label{eq:LAML}
\tilde{\mathcal{V}}(\bm \lambda) = \mathcal{L}(\hat{\bm \beta})+\frac{1}{2}\log|{\bf S}^{\bm \lambda}|_{+}-\frac{1}{2}\log|\bm {\mathcal{H}}|+\frac{M_{p}}{2}\log(2\pi),
\end{equation}
where  $M_{p}$ is the dimension of the null space of ${\bf S}^{\lambda}$, $|{\bf S}^{\lambda}|_{+}$ is the product of its positive eigenvalues, $\mathcal{L}(\bm \beta)$ is the r.h.s. of (\ref{eq:logPosterior}), $\hat{\bm{\beta}}$ is its maximiser and $\bm{\mathcal{H}}$ is its negative Hessian, evaluated at $\hat{\bm{\beta}}$. To ensure the positivity of $\bm \lambda$, we maximise (\ref{eq:LAML}) w.r.t. $\bm \rho$, where $\rho_g = \log(\lambda_g)$. We use a BFGS optimiser, which requires the gradient of the objective
\begin{equation} \label{eq:LAML_grad}
(\nabla_{\bm \rho} \tilde{\mathcal{V}})_g = \frac{\partial \tilde{\mathcal{V}}}{\partial \rho_g} = - \frac{\lambda_g}{2} \hat{\bm \beta}\ts {\bf S}_g \hat{\bm \beta} + \frac{1}{2}\frac{\partial \log|{\bf S}^{\bm \lambda}|_{+}}{\partial \rho_g} - \frac{1}{2}\frac{\partial \log|\bm {\mathcal{H}}|}{\partial \rho_g}.
\end{equation}
While computing the first two terms is straightforward, the third term requires implicit differentiation and third derivatives of the log likelihood w.r.t. $\bm \beta$, as explained in SM \ref{app:derivatives}.

\subsection{Interpreting the model via accumulated local effects } \label{sec:ale_effects}

Adopting a Bayesian framework to fit probabilistic additive stacking models allows us to use standard methods to quantify the uncertainty of the fitted regression coefficients, $\bm \beta$. In particular, we use an asymptotically justified approximation to $p(\bm \beta|\bm y, \bm \lambda)$ which consists of a Gaussian distribution, $N(\hat{\bm \beta}, {\bf V}_{\bm \beta})$, centered at the MAP estimator and with covariance matrix ${\bf V}_{\bm \beta} = -\bm {\mathcal{H}}^{-1}$. This posterior approximation ignores the uncertainty of the smoothing parameter estimates, which are considered fixed to the LAML maximiser. In principle, smoothing parameter uncertainty could be estimated via a Gaussian approximation to $p(\bm \lambda|\bm y)$ and then propagated forward to obtain an approximation to the unconditional posterior, $p(\bm \beta|\bm y)$. \cite{wood2016smoothing} provide formulae to do this, but we leave it for future work as approximating $p(\bm \lambda|\bm y)$ requires the Hessian of $\tilde{\mathcal{V}}$ w.r.t. $\bm \rho$, which is tedious to derive. Note that the smooth effects are linear combinations of the regression coefficients, hence it is straightforward to derive pointwise Bayesian credible intervals for the effects. See \cite{nychka1988bayesian} for an analysis of the asymptotic frequentist properties of such intervals.

Recall that we indicated with $f_{kj}(\bm x)$ the $j$-th effect appearing in $k$-th linear predictor, $\eta_k$. The experts' weights, $\alpha_1(\bm x), \dots, \alpha_K(\bm x)$, lay on the standard simplex in $\mathbb{R}^K$ and are related to the linear predictors via the parametrisation (\ref{eq:multinomParam}). The latter is non-linear, which can be problematic when interpreting the effects of the covariates on the weights. To see this, consider a model with three experts with $\eta_1=0$, $\eta_{2} = x\beta_1$, $\eta_{3} = x\beta_2$ and a single scalar covariate, $x$. If $\beta_1 > 0$, one might expect $\alpha_2(x)$ to increase with $x$ but, if $\beta_2 > \beta_1$, this is true only for $x < \log\{\beta_1/(\beta_2 - \beta_1)\}/\beta^2$. For larger values of $x$, $\alpha_2(x)$ decreases. This means that, even for simple models, plotting the effects specified in the linear predictors does not provide information regarding how the stacking weights behave. 
Given that additive stacking models are not black box models, appropriate visualisation of the covariates effects on the experts' weights is essential for model building and validation. Hence, we adopt the accumulated local effects (ALE) of \cite{apley2016visualizing} to better quantify and visualise the covariates effects. Here we describe how ALE are constructed and how their uncertainty can be quantified. Examples will be provided in Section \ref{sec:demandForecast}.

To simplify the notation, let us drop the index $k$ and indicate with $\alpha(\bm x)$ the weight of one of the experts. If we assume that $\alpha(\bm x)$ is differentiable w.r.t. the $j$-th covariate, then the main ALE effect of $x_j$ is 
\begin{equation}
\alpha_{j,ALE}(x) = \int_{x_{min, j}}^{x} \mathbb{E}_{\bm x_{\backslash j}}\{ \alpha^j(z_j, \bm x_{\backslash j}) | x_j = z_j \} d z_j - c,
\end{equation}
where $x$ is the value of $x_j$ at which we want to evaluate the effect, $c$ is a constant, $\bm x_{\backslash j}$ is $\bm x$ with the $j$-th element excluded, $\alpha^j = \partial \alpha / \partial x_j$ and $\mathbb{E}_{\bm x_{\backslash j}}\{ \cdot | x_j = z_j \}$ is a conditional expectation taken w.r.t. $p(\bm x_{\backslash j}|x_j = z_j)$. The choice $x_{min, j}$ is unimportant, as changing it simply shifts the effect vertically, hence in practice $x_{min, j}$ is set to just below the smallest observed value of $x_j$. As \cite{apley2016visualizing} explain, ALE effects avoid the extrapolation error which affects the partial dependence plots of \cite{friedman2001greedy} under correlated covariates.  

Uncentered ALE effects are defined by setting $c$ to zero and are estimated as follows. Let $x_{i,j}$ be the $i$-th observed value of $x_j$ and define a grid $z_{0,j}, \dots, z_{B,j}$ of values along $x_j$, such that $z_{0, j}$ and $z_{B,j}$ are the smallest and the largest observed values of $x_j$. Let $n_j(1), n_j(2), \dots, n_{j}(B)$ be the number of $x_{i,j}$'s falling in $[z_{0,j}, z_{1,j}), [z_{1,j}, z_{2,j}), \dots, [z_{B-1,j}, z_{B,j}]$. Indicate with $v_j(x) \in \{1, \dots, B\}$ the bin number in which an arbitrary value $x$ of $x_j$ belongs to and let $S_j(v)$ be the set such that, if ${\bm x}_i \in S_j(v)$, then $x_{i, j}$ belongs to the $v$-th bin. The uncentered ALE effect of $x_j$ is estimated by
\begin{equation} \label{eq:uncentALE}
\hat{\alpha}_{j,ALE}(x)=\sum_{v=1}^{v_{j}(x)}\frac{1}{n_{j}(v)}\sum_{\{i:x_{i,j}\in S_{j}(v)\}}[\alpha(z_{v,j},\bm{x}_{i,\backslash j})-\alpha(z_{v-1,j},\bm{x}_{i,\backslash j})], \;\; \text{with} \;\; \hat{\alpha}_{j,ALE}(z_{0, j}) = 0.
\end{equation}
Centered ALE effects, $\tilde{\alpha}_{j,ALE}(x)$, are defined by setting $c = \mathbb{E} \{ \alpha_{j,ALE}(x_j) \}$ and are estimated similarly. \cite{apley2016visualizing} consider black box models and quantify the uncertainty of the ALE effects via bootstrapping. In our context, it is possible to obtain uncertainty estimates more efficiently. In particular, we use the delta method to approximate the posterior variance of uncentered ALE effects using $\text{var}\{\hat{\alpha}_{j,ALE}(x)\} \approx \nabla_{\bm \beta}\ts \hat{\alpha}_{j,ALE}(x) {\bf V}_{\bm \beta} \nabla_{\bm \beta} \hat{\alpha}_{j,ALE}(x)$. SM \ref{app:ALEdeltaMethod} shows how to compute the gradient of the centered or uncentered ALE effects w.r.t. $\bm \beta$. It also covers the case where $x_j$ is a categorical variable.

Here we consider only the main ALE effects, but \cite{apley2016visualizing} define also higher order ALE effects and show that they lead to a functional ANOVA-like decomposition for $\alpha(\bm x)$. Let $\alpha_{ALE}(\bm x) = \mathbb{E}\{\alpha(\bm x)\} + \sum_j \tilde{\alpha}_{j,ALE}(x)$ be the leading term of such a decomposition. In Section \ref{sec:demandForecast} we report the fraction of variance of each stacking weight, $\alpha_k(\bm x)$, that is explained by the corresponding estimate of $\alpha_{ALE}(\bm x)$. The resulting $R^2$ coefficients quantify the importance of the main ALE effects, relative to the higher order interactions.

\section{Disaggregate electricity demand forecasting} \label{sec:demandForecast}

\subsection{Data and experts} \label{sec:dat_and_exp}

We consider the data set from the CER trial \citep{commission2012cer}, which contains electricity demand $y^c_i$, for $i = 1, \dots, N$, measured in kWh and at 30min resolution by smart meters at 2672 Irish households, $c = 1, \dots, C$. The data set covers the whole of 2010 and contains the following survey information about each household: $\text{sc}_c$ is a categorical variable indicating the occupation of the chief income earner; $\text{o}_c = 1$ if the customer owns the property and 0 otherwise; $\text{hw}_c = 1$ if the water heater is electric and 0 otherwise, and $\text{wg}_c$ indicates the number of white goods. We integrate the demand data with hourly temperatures, $T_1, \dots, T_N$, from the National Centers for Environmental Information (NCEI). The demand data was preprocessed to remove anomalous customers, such as those whose demand was always near-zero. We ended up with a data set of $2565$ customers. We removed special days (e.g., Christmas day) as well, because demand forecasts on such days typically require manual intervention. See SM \ref{app:detailsOnApplication} for further details on data preparation. Since many parts of the analysis are performed week by week, we enumerate consecutive weeks so that week 1 starts on Sunday the 3rd of January 2010.

We consider four experts, $M_1$ to $M_4$. In the following we outline the structure of the experts and we explain what features of the data each model is meant to capture. While $M_4$ is fitted to the whole data set, $M_1$ to $M_3$ model each household separately. Hence, we simplify the notation by omitting the index $c$ when describing $M_1$ to $M_3$. Under $M_1$ or $LastMonth$, the $i$-th predictive density is
\begin{equation} \label{eq:lastMonth}
p_{t_i}(y_i) = \frac{1}{30h} \sum_{k=1}^{30} \phi_0 \left( \frac{y_i - y_{i-48k}}{h} \right),    
\end{equation}
where $\phi_0$ is a Gaussian p.d.f., truncated below zero and re-normalised to take into account the fact that $y$ is non-negative, while $t_i \in \{1, \dots, 48\}$ is the time of day in half hours.  The bandwidth $h$ of this kernel density estimator is chosen via the rule of thumb of \cite{silverman1986density}. The strength of $LastMonth$ is that the distribution of $y$ is modelled non-parametrically and can change abruptly with $t_i$.  $M_2$ or $GaulssInd$ is a log-normal generalised additive model for location scale and shape \citep[GAMLSS, ][]{rigby2005generalized}. In particular, if we define $z_i = \log(y_i)$, then $z_i | \bm x_i \sim N(\mu_i, \sigma^2_i)$, where $\bm x_i$ is the $i$-th $d$-dimensional covariate vector and
$$
\mu_i
= \beta_0^\mu
+ \psi_1(D_i)
+ f_{1}(z_{i-48}) 
+ f_{2}(z_{i-336}) 
+ f_{3}(t_{i}) 
+ f_{4}(T^s_{i}),
$$
$$
\log (\sigma_i) 
=
\beta_0^\sigma
+ \psi_2(D_i)
+ f_{5}(t_{i}).
$$
Here $\beta_0^\mu$ and $\beta_0^\sigma$ are intercepts, $\psi_1(D_i)$ and $\psi_2(D_i)$ are parametric factor effects of the day of the week $D_i$, $f_1$ to $f_5$ are smooth effects and $T^s_i$ is the smoothed temperature, defined by $T_i^s = \alpha T^s_{i-1} + (1-\alpha) T_i$ with $\alpha = 0.9$. See SM \ref{app:detailsOnApplication} for more details on, for instance, the types of the spline bases used for the smooth effects. $GaulssInd$ is meant to capture smooth components of the daily individual profiles, shown in Figure \ref{fig:Irish_prof}e, as well as the temperature, calendar and autoregressive effects, which are typically used to model aggregated demand. We expect $GaulssInd$ to perform well on customers with regular consumption patterns, but to struggle with the abrupt changes shown in Figure \ref{fig:Irish_prof}f. The latter are meant to be captured by model $M_3$ or $Dynamic$, which is a log-normal GAM model, where $\mu_i$ is modelled only by a smooth effect of $t_i$, while $\sigma$ is considered constant. Let $j_i \in \{1, \dots, 365\}$ and $w_i \in \{1, \dots, 51\}$ be the day and the week to which $y_i$ belongs. While $GaulssInd$ is fitted to all the data from the weeks preceding $w_i$, $Dynamic$ is fitted only to the data from the three days preceding $j_i$, which makes it quicker to adapt. 

$M_4$ or $GaulssCommon$ is a log-normal GAMLSS model fitted to the demand of all customers jointly, rather than separately as in $M_1$ to $M_3$, hence we start to use again the index $c$. The mean and standard deviation models are
$$
\mu_{i}^c
= \beta_0^\mu
+ \psi_1(D_i)
+ \psi_2(\text{sc}_c)
+ \psi_3(\text{o}_c)
+ \psi_4(\text{hw}_c)
+ \psi_5(\text{wg}_c)
+ f_{1}(t_i) 
+ f_{2}(T^s_i)
+ f_3(\overline{y}_{i}^c)
,
$$
$$
\log (\sigma_i^c) = 
\beta_0^\sigma
+ \psi_6(D_i)
+ f_{4}(t_{i})
,
$$
where $\psi_2$ to $\psi_5$ are the parametric effects of the household specific binary or factor variables defined above, while $\overline{y}_{i}^c$ is the average consumption of customer $c$ up to the week $w_i - 1$. Being fitted to the data from all customers, $GaulssCommon$ is the only expert to capture the effects of the household specific survey variables. The smooth effects $f_1$ to $f_4$ are the same for all customers, hence $GaulssCommon$ is meant to capture the demand patterns that are shared across customers. Given that household demand dynamics are highly heterogeneous across customers, as shown in Figure \ref{fig:Irish_prof}e-f, this model produces highly biased prediction for most customers. However, it provides a baseline forecast useful to predict the demand of households with anomalous consumption patterns. Further, as we explain in Section \ref{sec:stackingEval}, we forecast demand using a rolling horizon, and the baseline forecast provided by this expert is especially useful at the beginning of the forecasting period, when only few weeks of data are available for each household. 

This section defined a set of experts designed to capture different features of household demand data. The next one proposes an additive stacking model designed to flexibly combine their predictions.

\subsection{Additive stacking model structure} \label{sec:stacking_model}

Let $p^c_k(y_i^c|\bm x^c_i)$ be the predictive density corresponding to household demand $y_i^c$ under model $M_k$, with $k = 1, \dots, 4$.
Recall that additive stacking forms a dynamic mixture of expert densities $p^c(y_i^c|\bm x^c_i) = \sum_k \alpha_k(\bm x^c_i) p_k^c(y_{i}^c|\bm x_i^c)$. While the first linear predictor $\eta^c_{1i}$ must be equal to zero for identifiability, the linear predictors for $M_2$ to $M_4$ are 
\begin{align} \label{eq:stackModelIrish}
\eta_{2i}^c = \, &
\beta_0^2 
+ \psi_1(D_i) 
+ \psi_2(\overline{y}_i^c) 
+ \psi_3(s_i^c) 
+ \psi_4(\gamma^{c1}_{2i}) 
+ \psi_5(\gamma^{c3}_{2i}) 
+ \psi_6(\gamma^{c7}_{2i}) 
+ \psi_7(\gamma^{c{j_i-1}}_{2i}) 
+ f_1(t_i)
+ f_2(j_i), \nonumber \\
\eta^c_{3i} = \, &
\beta_0^3
+ \psi_8(D_i)
+ \psi_9(\text{do}_i^c)
+ \psi_{10}(\gamma^{c1}_{3i}) 
+ \psi_{11}(\gamma^{c3}_{3i}) 
+ \psi_{12}(\gamma^{c7}_{3i}) 
+ \psi_{13}(\gamma^{c{j_i-1}}_{3i}) 
, \nonumber \\
\eta_{4i}^c = \, &
\beta_0^4
+ \psi_{14}(\gamma^{c1}_{4i}) 
+ \psi_{15}(\gamma^{c3}_{4i}) 
+ \psi_{16}(\gamma^{c7}_{4i}) 
+ \psi_{17}(\gamma^{c{j_i-1}}_{4i}) 
+ f_3(j_i)
+ f_4(t_i),
\end{align}
where $\beta_0^2$, $\beta_0^3$ and $\beta_0^4$ are intercepts, $\psi_1(D_i)$, $\psi_8(D_i)$ and $\psi_9(\text{do}_i^c)$ are parametric factor effects, while all the remaining $\psi$'s are linear effects of continuous covariates. We now define each covariate and we explain why we use it within model (\ref{eq:stackModelIrish}).

Variables $\overline{y}_i^c$ and $s_{i}^c$ are the mean and standard deviation of the consumption of customer $c$ up to the week $w_i - 1$. We add their effects to $\eta_{2i}^c$, because $GaulssInd$ is the most complex by-household expert in the mixture and we expect it to do well on customers with rich consumption dynamics, which generally have high values of $\overline{y}_i^c$ and $s_{i}^c$. Figure \ref{fig:Irish_prof}a-d show that time of day, $t_i$, is a strong driver on demand dynamics, hence we add its effect to two of the linear predictors. We do not add it to the linear predictor of the $Dynamic$ expert because, as we explain below, we expect that its weight should depend on how household behaviour changed during the last few days, rather than on the daily demand pattern. We add the effect of the time of year $j_i$ to $\eta_{4i}^c$ because, as explained above, $GaulssCommon$ provides a baseline prediction which we expect to become less useful as more data becomes available. The categorical variable $\text{do}_i^c \in \{0, 1, 2, \geq 3\}$ indicates for how many days customer $c$ has been out of home before day $j_i$. Customers are considered to be out of home on a given day if the range of their consumption on that day is below 0.5 kWh. The effect of $\text{do}_i^c$ appears in $\eta^c_{3i}$ because the $Dynamic$ expert is the meant to react quickly to sudden changes in demand which occur, for example, when the household goes on holiday. We also used the variables
\begin{equation} \label{eq:relativPerf}
\gamma^{cu}_{ki} = \frac{\{\prod_{l=1}^u p_k^c(y^c_{i-48l})\}^{1/u}}{\sum_{m = 1}^4 \{\prod_{l=1}^u p_m^c(y^c_{i-48l})\}^{1/u}} = \frac{\exp \frac{1}{u} \sum_{l = 1}^{u} \log p_k^c(y^c_{i-48l})  }{\sum_{m=1}^4 \exp \frac{1}{u} \sum_{l = 1}^{u} \log p_m^c(y^c_{i-48l}) } \in (0, 1),
\end{equation}
where we omitted the dependence of $p_k^c$ on $\bm x_i^c$ for convenience. These variables capture the relative predictive performance of model $M_k$ on customer $c$ and at the same time, $t_i$, of the $u$ days preceding $j_i$. For example, $\gamma^{c1}_{ki}$ is ratio between the predictive density under $M_k$, $p_k^c(y^c_{i-48})$, and the average of the predictive densities under $M_1$ to $M_4$. Therefore, $\gamma^{c1}_{ki} \approx 1$ indicates that $M_k$ provided a much better probabilistic prediction of $y^c_{i-48}$, relative to the other experts. The predictive performance of $M_k$ on  $y_{i-48}^c$ should provide information on how well it will predict $y_i^c$. Hence, we use the linear effect of $\gamma^{c1}_{ki}$ in (\ref{eq:stackModelIrish}) to let the weight of the $k$-th expert vary with the performance of $M_k$ at the same time of the previous day. The interpretation of $\gamma^{cu}_{ki}$ with $u = 3, 7$ or $j_i-1$ is similar. In particular, they use the geometric means, over several lags, of the predictive densities of $M_k$ and of the other experts, to capture the relative predictive performance of $M_k$ over several days preceding $j_i$. If past relative performance is positively correlated with future performance, then we should expect the linear effects of the $\gamma^{cu}_{ki}$'s to be positive. Note that the past performance of each expert could have been quantified in a number of ways. Our choice is based on the relation between (\ref{eq:relativPerf}) and the exponentially weighted average forecaster (EWA), which is a simple expert aggregation strategy \cite[for an introduction see, e.g.,][]{cesa2006prediction}. In fact, $\gamma^{cj_i-1}_{ki}$ is the weight that would be attributed to $M_k$ by an EWA forecaster based on the log-loss ($\text{Loss}_i = -\log p_k^c(y^c_{i})$) and with learning rate equal to $1/(j_i-1)$. Hence, we are using the EWA weights at different lags as covariates in the additive stacking model.

In the next Section we evaluate the predictive performance of the experts and of model (\ref{eq:stackModelIrish}) on the Irish household demand data. We will also examine and interpret the fitted stacking model via ALE effects plots.

\subsection{Stacking model evaluation and visualisation}
\label{sec:stackingEval}

Recall that stacking models must be fitted to out-of-sample data, because using the same data to fit the experts and the stacking model would lead to overfitting. Here we fit the models and evaluate their predictive performance using the following procedure. We use the data from weeks 1-5 to fit the experts $GaulssInd$ and $GaulssCommon$, which can then provide an half-hourly probabilistic forecast for the whole of week 6. We store this forecast, then we use the data from weeks 1-6 to fit the experts and we produce a forecast for week 7. By iterating this fitting and forecasting procedure until week 51, we obtain out-of-sample probabilistic forecasts from these two models for weeks 6-51. Note that we do not use the data for week 52 because electricity demand during this period is atypical due to holidays (in an operational setting, the demand forecast for week 52 requires manual adjustments). For the $LastMonth$ and $Dynamic$ experts we follow a similar procedure, but we updated the models more frequently. In particular, let $j$ be the index of the first day of week 6. We fit the $Dynamic$ model to data from days $j-3$ to $j-1$ and we use it to produce a half-hourly probabilistic forecast for the whole of day $j$. We do the same under the $LastMonth$ model, but using data from days $j_i-30$ to $j_i-1$ within (\ref{eq:lastMonth}). Thus, for weeks 6-51, we have out-of-sample probabilistic predictions from all experts, which can be used to fit the stacking model.

\begin{figure} 
\centering
\includegraphics[width=\textwidth]{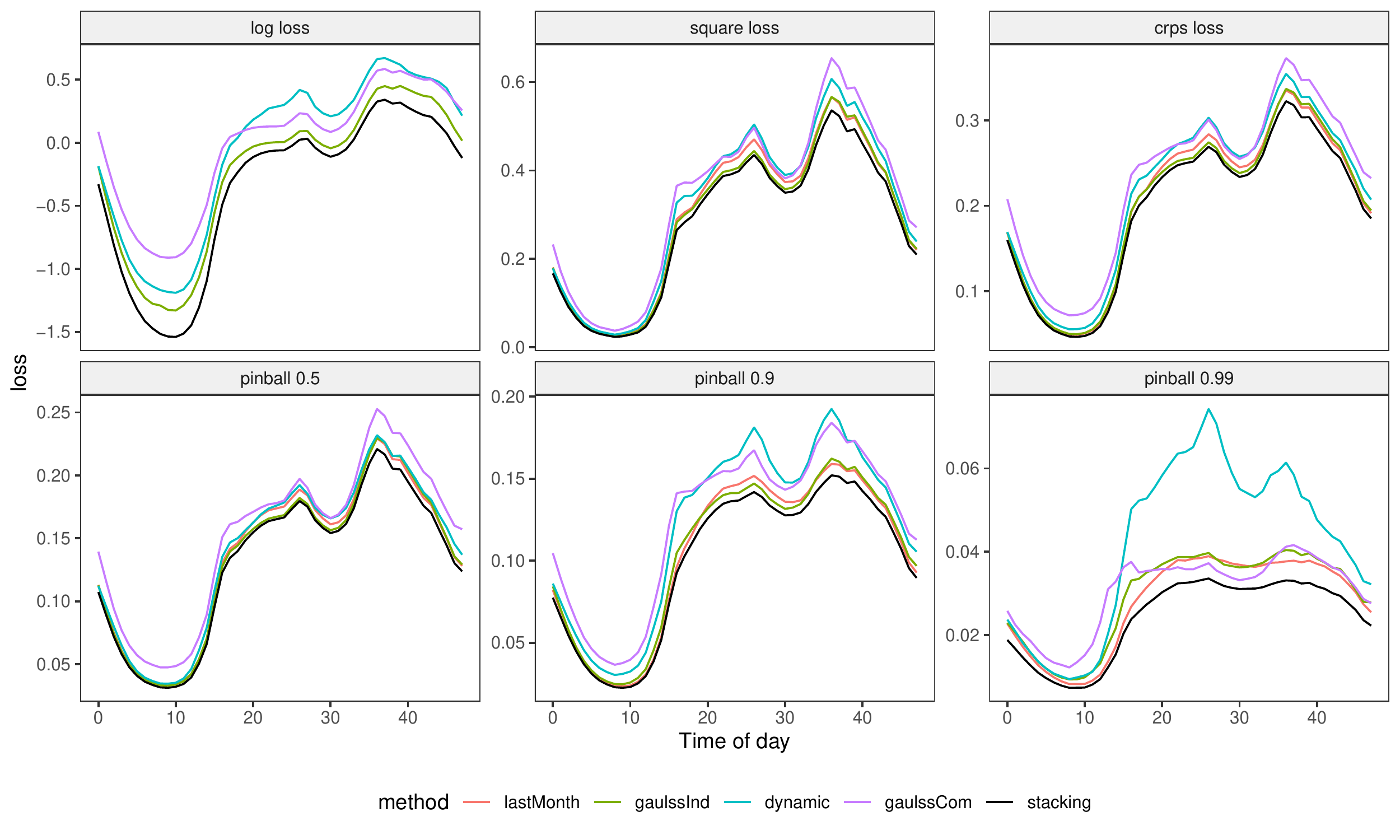} 
\cprotect\caption{Mean of predictive losses along the time of the day.
Each panel corresponds to a loss function, and each line corresponds to a single method being evaluated. 
Each line is obtained by averaging the loss function over all customers and days of the year of weeks from 10 to 51, at each time of the day.
Note that the log-loss of \textit{LastMonth} is not shown since it is much larger than the log loss of the other experts.}
\label{fig:predictive_scores}
\end{figure}

Before fitting the stacking model to all the available out-of-sample data, we compare its predictive performance with that of the experts. In particular, we fit the stacking model (\ref{eq:stackModelIrish}) to the data from weeks 6-9 and we use it to produce predictions for week 10. In the next step, we fit it using the data from weeks 6-10 and we predict the demand on week 11. By iterating this, we obtain stacked predictions for weeks 10-51, which can be compared with those produced by the experts. Figure \ref{fig:predictive_scores} shows the results of such a comparison. In particular, we quantify the predictive performance of each model using several loss functions and we plot the average losses as functions of the time of day. We consider the log-loss, which is simply the negative log-likelihood evaluated on the test data, the square loss, the continuous ranked probability score (CRPS) and the pinball loss. Note that the losses achieved by additive stacking are strictly lower than those of the experts, at any time of day and under any loss type. This is remarkable, as model (\ref{eq:stackModelIrish}) was fitted via likelihood-based MAP and LAML methods which are directly related to the log-loss, but not to the other losses. Under any method and loss, the predictive performance is better at night, when the demand is low and stable, than during the two daily peaks. As explained in  \cite{koenker1978regression}, the pinball loss is parametrised by $\tau \in (0, 1)$, and it is minimised by the conditional quantile $Q_\tau(y|\bm x)$ corresponding to probability level $\tau$. We evaluate the pinball loss at three levels of $\tau$ and additive stacking achieves larger improvements, relative to the experts, on the highest quantiles. This suggests that stacking is doing a better job at predicting the daily demand spikes. The loss of $LastMonth$ is missing from the log-loss plot in Figure \ref{fig:predictive_scores}, because this expert performs very poorly under this loss, as detailed in Table \ref{tab:predictive}. The poor performance of $LastMonth$ on this loss is due to the fact that this expert is based on a thin tailed mixture of Gaussian densities (\ref{eq:lastMonth}), which generates large losses on outlying demand observations. However, $LastMonth$ is more competitive on the other losses and in the following we illustrate that, surprisingly, it is often the expert to which additive stacking attributes the largest weight. 

\begin{table}
\begin{tabular}{r|c|c|c|c|c|c}
                & Log-loss  & CRPS & Square loss & Pinball 0.5 & Pinball 0.9 & Pinball 0.99          \\ \hline
LastMonth            & 28.812  & 0.204  & 0.298  & 0.136 & 0.105 & 0.028 \\
GaulssInd           & -0.244   & 0.203  & 0.291  & 0.134 & 0.105 & 0.030 \\
Dynamic             & -0.059  & 0.216   & 0.317  & 0.138 & 0.121 & 0.042 \\
GaulssCommon          & -0.019 & 0.230  & 0.335  & 0.151 & 0.123 & 0.031 \\
Stacking              & \textbf{-0.376} & \textbf{0.195}  & \textbf{0.279}  & \textbf{0.130} & \textbf{0.100} & \textbf{0.024}         
\end{tabular}
\caption{Mean predictive losses of each model. The lowest loss in each category is \textbf{bold}.}
\label{tab:predictive}
\end{table}

\begin{figure} 
\centering
\includegraphics[width=\textwidth]{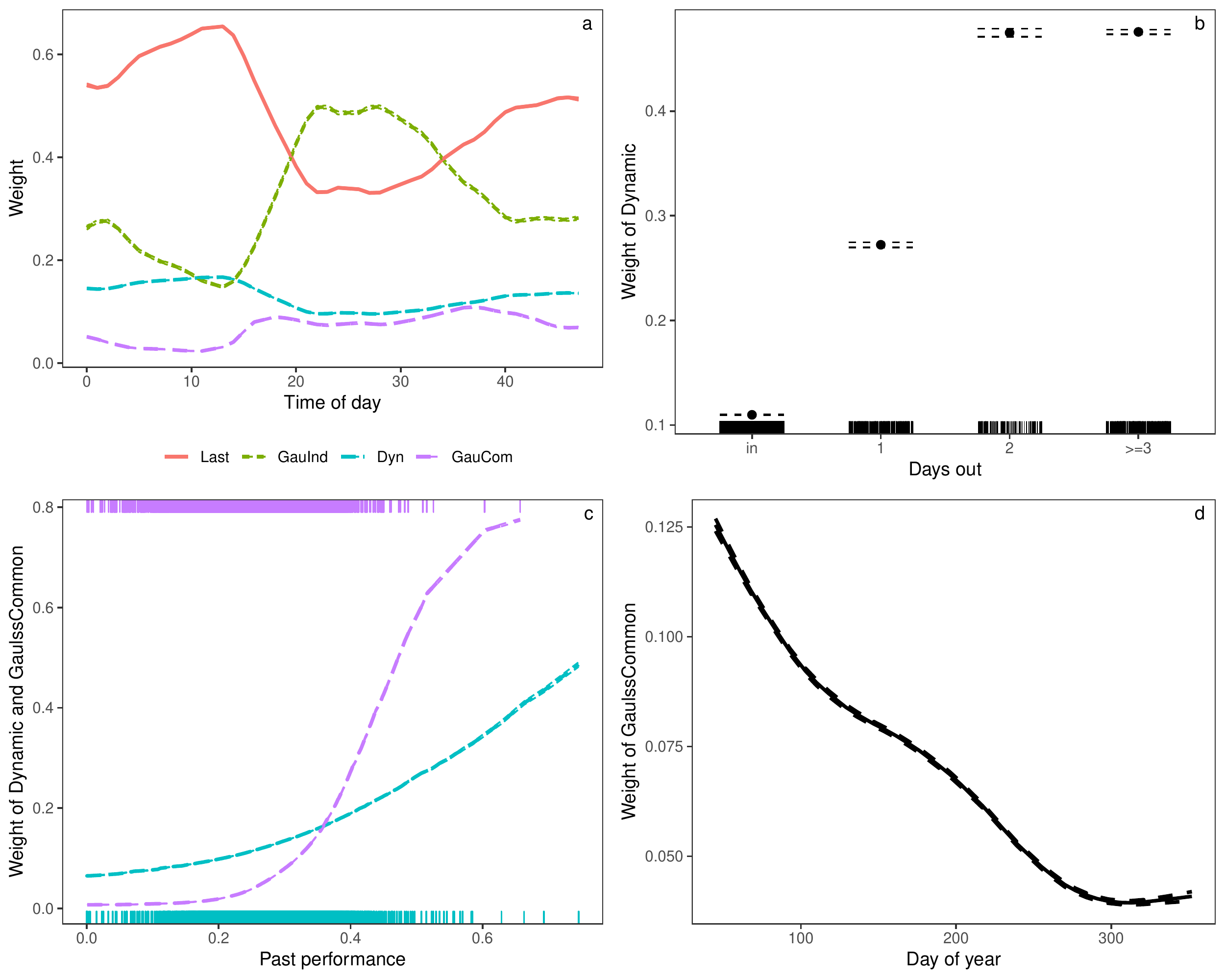} 
\cprotect\caption{Centered ALE effects on the stacking weights of a) each expert, b) $Dynamic$, c) $Dynamic$ and $GaulssCommon$ and d) $GaulssCommon$. The $x$-axis in c) represents the $\gamma^{cj_i-1}_{ki}$ covariate, defined by (\ref{eq:relativPerf}). The plots have been shifted vertically by adding the average weight of each expert. The credible intervals for some of the effects are too narrow to be visible, which is not surprising given that the stacking model was fitted to over $30 \times 10^6$ observations.}
\label{fig:ALE_irish}
\end{figure}

Figure \ref{fig:ALE_irish} shows several effects plots, obtained by using the ALE methods described in Section \ref{sec:ale_effects} on an additive stacking model fitted to data from weeks 10-51. The plots have been produced using the \verb|mgcViz| \verb|R| package \citep{fasiolo2018scalable} and show centered ALE effects, which have been shifted vertically by adding the average weight of each expert. Figure \ref{fig:ALE_irish}a shows that $LastMonth$ is the expert with the largest average weight, but $GaulssInd$ becomes the dominant model during the key working hours. The stacking model attributes a large weight to $LastMonth$ during the night, when the demand is consistently low and can be predicted effectively using past observations, as done by (\ref{eq:lastMonth}). During daytime, demand dynamics are more complex and depend on factors, such as the day of the week, which are captured by $GaulssInd$. $Dynamic$ has on average a low weight, but Figure \ref{fig:ALE_irish}b shows that its weight depends strongly on the out-of-home, $\text{do}_i^c$, variable. In particular, recall that $Dynamic$ uses only data from the last three days, which makes it quick to adjust when a customer leaves home and the demand suddenly drops. Figure \ref{fig:ALE_irish}c shows the ALE effect of the $\gamma^{cj_i-1}_{ki}$ variables corresponding to the $Dynamic$ and $GaulssCommon$ experts, on their own weights. Recall that $\gamma^{cj_i-1}_{ki}$ measures the predictive performance of the $k$-th method using all the past data from customer $c$. The plot shows that the weight of $Dynamic$ can reach around 0.4 on some customers. These are customers who frequently vacate their homes. The by-customer past performance has a strong effect on the weight of $GaulssCommon$. In fact, while Figure \ref{fig:ALE_irish}a shows that its average weight is below 0.1, Figure \ref{fig:ALE_irish}c illustrates that it can be the dominant expert on some customers. As explained in Section \ref{sec:dat_and_exp}, $GaulssCommon$ provides a baseline forecast, useful for household with irregular demand patterns. Further, the baseline forecast provided by $GaulssCommon$ is more useful when little customer specific data is available. In fact, Figure \ref{fig:ALE_irish}d shows that the stacking model reduces its weight as more weeks of data become available. Note that the marginal or first order ALE effects explain a large fraction of the variance of the experts' weights. In particular, for models $M_1$ to $M_4$, we get $R^2 = 0.81, 0.92, 0.88$ and $0.69$, hence higher order interactions are strongest for the weights of the $GaulssCommon$ model which, however, has the smallest weight in the mixture. 

In this section we demonstrated that additive stacking improves upon the predictions provided by the experts, under several loss functions. We also showed how the effects of the covariates on the stacking weights can be visualised using ALE plots. While we focused on a subset of such plots, ALE plots for all possible expert/covariate pairs can be found in SM \ref{app:detailsOnApplication}. In the next section we show how the household specific probabilistic forecast generated by additive stacking can be used within the home battery scheduling application mentioned in Section \ref{sec:introduction}.

\subsection{Home battery scheduling} \label{sec:peakShave}

Let $\bm y^c = \{y^c_1, \dots, y^c_{48}\}$ be the demand of household $c$ on a given day and consider a future scenario where the electricity bill of each households is determined via the daily-max tariff
\begin{equation} 
\label{eq:peak_tariff}
    \text{B}(\bm y^c) = P \sum_{t=1}^{48} y^c_t + P \cdot 48 \max_{t=1,\dots,48} y_t^c,
\end{equation}
where $t$ is the time of day and $P = 0.16$\euro/kWh, which is around one third less than the current unit price in Ireland, according to Eurostat. 
Hence, a customer pays price $P$ for each  kWh of total consumption and a further $48 P$ for each unit of peak consumption. Other choices of $P$ and of the peak price multiplier are obviously possible. The chosen setting is motivated by simplicity: one unit of peak consumption is as costly as one unit consumed at each time of the day. To our best knowledge, customer specific peak tariffs are not yet available to residential customers. For example, in the SmartHours program offered by OGE Energy in Oklahoma, the same price signal is sent to all customers one day ahead, with prices ranging from $5$\textcentoldstyle \textcolor{white}{o}to $41$\textcentoldstyle \textcolor{white}{o}for critical hours. Under high home battery penetration, such a tariff might lead to undesirable battery coordination across households, as suggested by \cite{pimm2018time}. Ontario's power system operator created an Industrial Conservation Initiative where customers are charged based on their contribution to major aggregate demand peaks. The tariff is currently available only to industrial customers, but it might be a suitable alternative to the one proposed here to avoid battery induced peaks.  

Assume that each household is equipped with a home battery of usable capacity $soc_{max} = 6.9$kWh, maximum charge/discharge rate $\delta_{max} = 3.75$kW, efficiency $\epsilon = 0.97$ and a useful life of $n_{cy} = 8000$ complete charging cycles. These are the specifications of a Mercedes-Benz energy storage home battery composed of three 2.5kWh modules. Let $\bm \delta^c = \{\delta_1^c, \dots, \delta_{48}^c\}$ be the daily  charge/discharge schedule for battery $c$ and assume that $\bm \delta^c$ is planned one day ahead, by minimising the expected daily cost to the customer. In particular, if $soc_t$ is the state of charge of the battery at time $t$, $\mathbbm{1}$ is the indicator function and $A$ the price of the battery, then the optimal schedule is 
\begin{equation}
\label{eq:objective_fun}
    {\bm \delta}^c = \underset{\bm \delta}{\text{argmin}} \, \left[
    \mathbb{E} \left \lbrace \text{B}(\bm y^c + \bm \delta) \right \rbrace
    + \sum_{t=1}^{48} \mathbbm{1} (\delta_t < 0) \lvert \delta_t \rvert \frac{A}{n_{cy} soc_{max}} \right],
\end{equation}
with $|\delta_t| \leq \delta_{max}$ and 
$$
\label{eq:efficiency}
\text{soc}_t = 
\begin{cases}
\min(\text{soc}_{t - 1} + \delta_t \epsilon, soc_{max}) & \text{if } \; \delta_t \geq 0 \\
\max(\text{soc}_{t - 1} + \delta_t / \epsilon, 0) & \text{if } \;  \delta_t < 0.
\end{cases}
$$
for $t = 1, \dots, 48$. We do not force $y_t + \delta_t$ to be positive, hence the battery can potentially sell power to the grid at price $P$, but we impose the additional constraint $s_0 = s_{48} = 0.1 soc_{max}$ to prevent the battery from being completely discharged at the end of each day.  

Note that the loss (\ref{eq:objective_fun}) is the sum of the expected electricity bill and the cost of battery usage. Given that we are considering a one day-ahead planning horizon, we assume that the expectation in (\ref{eq:objective_fun}) is conditional on the information available on the previous day. While the one-day-ahead household demand distribution is unknown, it can be estimated as described in Section \ref{sec:stackingEval}. In particular, we use the probabilistic day-ahead forecasts produced by the experts and the stacking model to estimate the expected value in (\ref{eq:objective_fun}), for each customer and each day. For each expert, we estimate the expected daily bill by simulating $10^3$ samples from the estimated day-ahead demand distribution. The estimated loss is minimised separately for each customer and day, using a constrained BFGS algorithm. 

\begin{figure} 
\centering
\includegraphics[width=\textwidth]{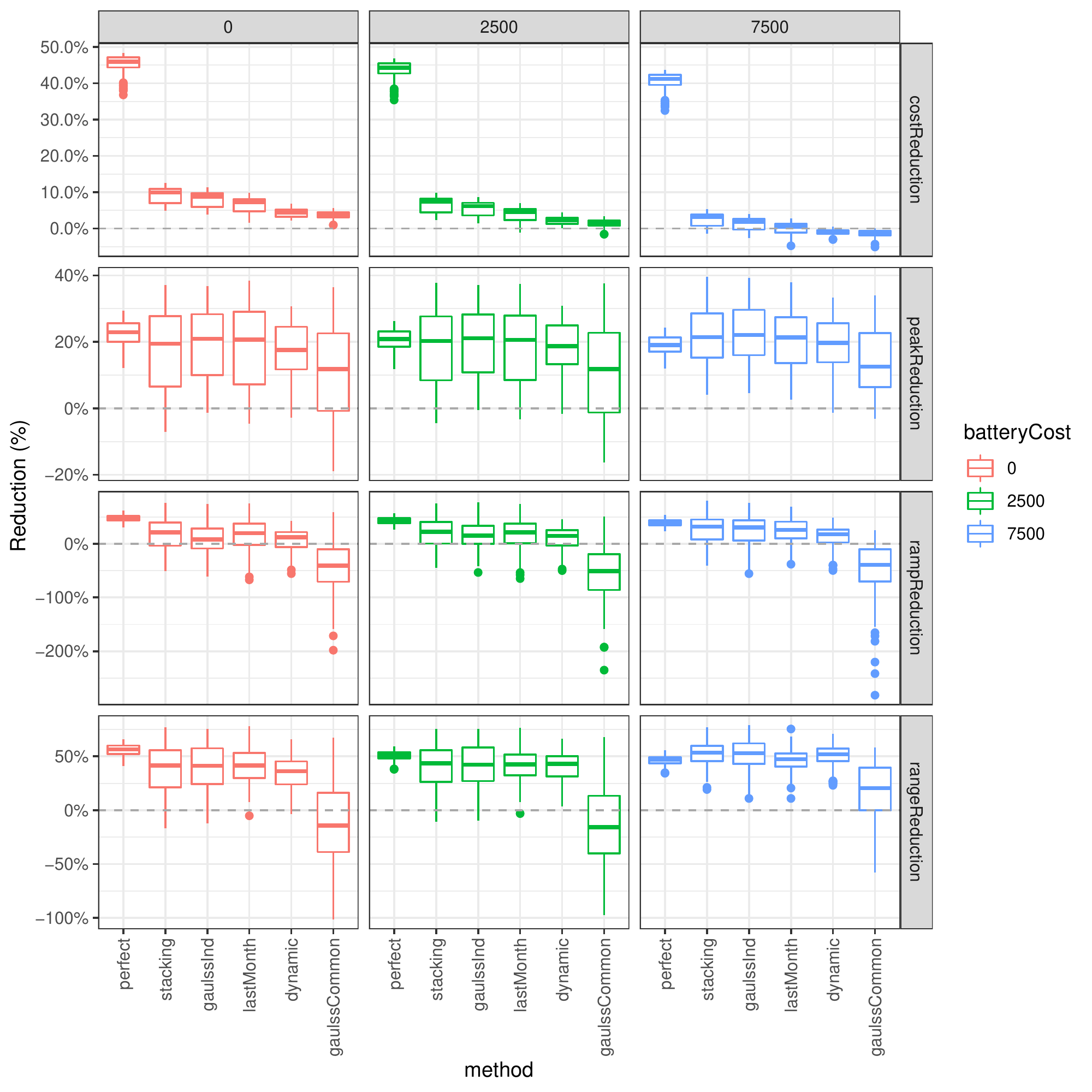} 
\cprotect\caption{Boxplots showing the distribution, over all days of the year between weeks 10 and 51, of the percentage reduction of the aggregate daily electricity cost, peak of the aggregate electricity demand and ramp, relative to leaving the batteries idle. The results are provided for three different battery cost scenarios. Under the `perfect' expert, the household demand is assumed to be known one day in advance.}
\label{fig:reduction}
\end{figure}

We consider three possible prices, $A =$ 7500, 2500 and 0\euro, for the home battery. The first is close to current prices, while the other two prices correspond to hypothetical scenarios where the batteries are subsidised or cheaper to produce. The plots in the first row of Figure \ref{fig:reduction} show the daily cost reduction, aggregated over all customers and relative to leaving the batteries idle, achieved by minimising the loss estimated under each model. To provide an upper bound on the potential savings, we include in the comparison a `perfect' expert, under which the demand is known one day in advance. Note that additive stacking leads to more savings, relative to the other experts, under each battery cost scenario. This is not surprising, as additive stacking produces more accurate probabilistic forecasts than the individual experts (see Section \ref{sec:stackingEval}), thus leading to better estimates of the loss (\ref{eq:objective_fun}). However, note the large gap between the cost reduction achieved by stacking and the upper bound provided by the perfect expert, as well as the fact that the upper bound is less sensitive to the battery cost than the reduction achieved by stacking or any of the experts. Both issues are related to the adoption of a daily-max tariff. In particular, household level demand is characterised by large, hard to predict, spikes (see Figure \ref{fig:Irish_prof}) which are heavily penalised by tariff (\ref{eq:peak_tariff}). Hence, further improving the cost reduction achieved by additive stacking would require improved predictions of the daily demand spikes of individual households. But this would probably require intra-day, household specific, information, while here we are considering a day-ahead planning horizon.

It is interesting to verify what is the effect of battery scheduling under the daily-max tariff on the daily aggregate electricity demand profile. In particular, \cite{pimm2018time} present a simulation based study focused on the effect of home batteries on the aggregate demand profile and show that, under standard time-of-use tariffs, there might be little or no reduction in the daily peak demand. They show that peak consumption might even increase, if all batteries start charging simultaneously at the beginning of the off-peak price band. They suggest that a tariff based on the individual daily peak demand might lead to better peak shaving results, which is something that we can verify here. The plots in the second to last row of Figure \ref{fig:reduction} show the reduction, relative to no battery usage, in the daily peak, range and ramp of aggregate demand, achieved under each model and battery cost scenario. The range is simply the difference between the daily maximum and minimum of the aggregate demand, while the ramp is the largest daily absolute difference between the aggregate demand at two consecutive time points. Figure \ref{fig:reduction} suggests that all experts lead to some reduction in peak demand, particularly under the full battery cost scenario. This is because taking the full cost of the battery into account makes it less likely that many batteries will be charged at night, when the demand is low across most households. In fact, Figure \ref{fig:peak_plot}d shows that, under a fully subsidised scenario, simultaneous battery charging leads to demand peaks during the night. The problem is particularly severe when the forecasts provided by the $GaulssCommon$ expert are used, because similar daily demand profile forecasts are used to optimise all the batteries. Indeed, the concurrent action of all batteries under the $GaulssCommon$ forecast has a destabilising effect on the system. In particular, the last two rows of Figure \ref{fig:reduction} show that this expert often leads to an increase in the daily demand ramp and range, thus making the system more difficult to manage.

Under additive stacking and the other three experts, the problems just described are less severe because the forecasts are better tailored to each household, leading to less correlated battery schedules across households. The performance of additive stacking in terms of the peak, ramp and range reduction is roughly comparable to that of these three experts. We should not expect stacking to do better than the experts on these scores, as it did for cost reduction, because we are looking at side effects of the individually optimised battery schedule on the aggregate demand. Indeed, even the `perfect' expert does not beat the other models in terms of peak reduction, when $A = 7500$\euro. However, under a perfect forecast, the percentage of peak, ramp and range reduction have very low variability and are less sensitive to the cost of the battery than for the other forecasts. As for the gap between the upper bound and additive stacking for price reduction, this is due to the difficulty of predicting the time and size of the individual demand peaks, one day in advance. Under the perfect forecast, the time and size of the peaks are known and can be used to generate highly specific battery schedules. The resulting aggregate demand profile is a flatter version of the original profile, as shown in Figure \ref{fig:peak_plot}d. Given that the individual household demand peaks are difficult to predict one day ahead, additive stacking and the experts produce smoother daily household demand forecasts, which lead to more correlated battery schedules. The result is an aggregate demand profile which, while being flatter than the original profile, has a shallow demand peak at night (see Figure \ref{fig:peak_plot}d).

The results presented in this section suggest that optimising home batteries schedules, separately for each household and under a daily-max schedule, could lead to a flatter aggregate daily profile, which is a desirable outcome from an electrical grid management point of view \citep{pimm2018time}. However, battery schedule optimisation must be based on probabilistic forecasts that are specific to each household, such as those generated by additive stacking. 
In terms of cost reduction, getting closer to the upper bound provided by the `perfect' expert would require more accurate predictions of the daily household demand peaks. We doubt that such predictions could be obtained one day ahead. Instead, we feel that they would require the adoption of a shorter (i.e. intra-day) forecasting horizon and possibly the use of additional, household specific, covariates.

\section{Conclusion} \label{sec:conclusion}

We focused on probabilistic electricity demand forecasting at the individual household level via additive stacking. The stacking ensemble members are predictive densities and the ensemble weights are allowed to vary with the covariates by adopting an additive model structure. In particular, the weights can be modelled via fixed, random or smooth effects based on spline basis expansion. The household demand data set considered here includes over 30 million observations, hence we fitted the stacking model using fast direct MAP and LAML methods for regression coefficients estimation and smoothing parameters selection. 

To capture different features of household demand, we developed a set of four heterogeneous probabilistic experts. While the experts were fitted to individual household data, the additive stacking model estimated the experts' weights using data from all households. This allowed it to borrow information across households, thus reducing the variance, while accommodating for the heterogeneity of household demand. The results are encouraging because, while being fitted using Bayesian likelihood based methods, the stacking model beats all experts under several loss functions. In addition, the home battery scheduling results show that, using the probabilistic demand forecasts produced by additive stacking for home battery optimisation under a daily-max tariff, leads to larger cost savings than under any of the experts. Further, planning the charge/discharge schedule of each battery under individually tailored demand forecasts has the desirable side effect of making the daily aggregate demand profile flatter. 

\section*{Acknowledgements}
This work was partially funded by EPSRC grant EP/N509619/1 and by EDF. The authors are thankful to Jethro Browell and Stephen Haben for helpful discussions on the use of disaggregate forecasts for grid management. 

%
%

\bibliographystyle{chicago}
\bibliography{biblio.bib}

\pagebreak
\clearpage
\begin{center}
\textbf{\large Supplementary material to ``Additive stacking for disaggregate electricity demand forecasting''\\}
\vspace{2mm}
Christian Capezza, Biagio Palumbo, Yannig Goude, Simon N. Wood and Matteo Fasiolo
\end{center}
\setcounter{section}{0}
\setcounter{equation}{0}
\setcounter{figure}{0}
\setcounter{table}{0}
\setcounter{page}{1}
\makeatletter
\renewcommand{\theequation}{S\arabic{equation}}
\renewcommand{\thefigure}{S\arabic{figure}}
\renewcommand{\thesection}{S\arabic{section}}
\renewcommand{\bibnumfmt}[1]{[S#1]}
\renewcommand{\citenumfont}[1]{S#1}

\section{Derivatives of the additive stacking log-likelihood w.r.t. $\bm \beta$} \label{app:derivatives}

In this section, we provide the derivatives needed to fit probabilistic additive stacking models using the methods described in the main text. Consider a function $f$ of the $N$-dimensional vectors $\bm \eta_1, \dots, \bm \eta_K$. 
We indicate with $f^{\bm \eta_k}$, $f^{\bm \eta_k, \bm \eta_j}$ and $f^{\bm \eta_k, \bm \eta_j, \bm \eta_m}$ the vectors with
$i$-th elements  
$$
f^{\eta_{ki}} = \frac{\partial f}{\partial \eta_{ki}}, \quad
f^{\eta_{ki} \eta_{ji}} = \frac{\partial^2 f}{ \partial \eta_{ki} \partial \eta_{ji}} \quad \text{and} \quad
f^{\eta_{ki} \eta_{ji} \eta_{mi}} = \frac{\partial^3 f} {\partial \eta_{ki} \partial \eta_{ji} \eta_{mi}},
$$ 
where $\eta_{ki}$ indicates the $i$-th elements of $\bm \eta_k$.
Each $\bm \eta_k$ is a function of a corresponding $p_k$-dimensional vector $\bm \beta^k$ and we indicate the Jacobian $\nabla_{\bm \beta^k}\ts \bm \eta_k$ with $\bm \eta_k^{\bm \beta^k}$. For the derivatives of $f$ w.r.t. the elements of the $\bm \beta^k$'s, we use the following compact notation
$$
f^{\beta_r^k} = \frac{\partial f}{\partial \beta^k_r}, \quad
f^{\beta_r^k \beta_s^j} = \frac{\partial^2 f}{ \partial \beta_r^k \partial \beta_s^j} \quad \text{and} \quad
f^{\beta_r^k \beta_s^j \beta_t^m} = \frac{\partial^3 f} {\partial \beta_r^k \partial \beta_s^j \beta_t^m},
$$ 
where $\beta_r^k$ indicates the $r$-th elements of $\bm \beta^k$. Finally, we denote with $f^{\bm \beta^k} = \nabla_{\bm \beta^k} f$ the gradient of $f$ w.r.t. $\bm \beta^k$ and with $f^{\bm \beta^k\bm \beta^j} = \nabla_{\bm \beta^j}\ts \nabla_{\bm \beta^k} f$ a matrix of second derivatives. 

\subsection{Gradient and Hessian of the additive stacking log-likelihood w.r.t. $\bm \beta$}

In this section we provide the gradient and Hessian of the penalised posterior log-density w.r.t. $\bm \beta$, which are required for maximisation using Newton's algorithm, for fixed smoothing parameters $\bm \lambda$. To simplify the notation, let us define $\mathcal L (\bm \beta) = \log p(\bm \beta|\bm y, \bm \lambda)$ and recall that
$$
    \mathcal L(\bm \beta) = \sum_{i=1}^N l_i - \frac{1}{2} \sum_{g=1}^G \lambda_g \bm \beta^\top \mathbf S_g \bm \beta,
$$
where $l_i = \log \sum_{k=1}^K \alpha_{ki} (\bm \beta) p_k (y_i | \bm x_i)$ is the log-likelihood term relative to the $i$-th observation.
The gradient of the log-posterior w.r.t $\bm \beta$ is
$$
    \mathcal L^{\bm \beta} (\bm \beta) = \sum_{i=1}^N l_i^{\bm \beta} - \sum_{g=1}^G \lambda_g \mathbf S_g \bm \beta,
$$
while the Hessian of the penalised log-likelihood w.r.t $\bm \beta$ is
$$
    \mathcal L^{\bm \beta \bm \beta} (\bm \beta)
    =
    \sum_{i=1}^N l_i^{\bm \beta \bm \beta} - \sum_{g=1}^G \lambda_g \mathbf S_g,
$$
Therefore, we need to calculate $l^{\bm \beta}$ and $l^{\bm \beta \bm \beta}$.
First of all, we arrange the regression coefficients as $\bm \beta = ({\bm \beta^2}^\top, \dots, {\bm \beta^K}^\top)^\top$, where $\bm \beta^k$ is the vector of regression coefficients specific of the $k$-th linear predictor (recall that the first linear predictor is set to the zero vector for identifiability).
The $k$-th linear predictor is $\bm \eta_{k} = \mathbf X^k \bm \beta^k$, where $\mathbf X^k$ is an $N \times p_k$ model matrix.
Given that $\bm \eta_{k} ^ {\bm \beta^k} = \mathbf X^k$, we can use the chain rule to calculate the derivatives of the log-likelihood with respect to the $\bm \beta^k$'s once we have derivatives of the log-likelihood with respect to the linear predictors. 
We can express $l_i$ as function of the linear predictors
$$
l_i =  \log \sum_{k=1}^K \exp ({\eta_{ki}} + \log p_k (y_i | \bm x_i)) - \log \sum_{k=1}^K \exp{\eta_{ki}}.
$$
Then, we have
$$
l_i^{\eta_{ki}}
= \frac{\exp ({\eta_{ki}} + \log p_k (y_i | \bm x_i))}{\sum_{h=1}^K \exp ({\eta_{hi}} + \log p_h (y_i | \bm x_i))} - \frac{\exp \eta_{ki}}{\sum_{h=1}^K \exp \eta_{hi}}.
$$
By defining
$$
w_{ki} = \frac{\exp ({\eta_{ki}} + \log p_k (y_i | \bm x_i))}{\sum_{h=1}^K \exp ({\eta_{hi}} + \log p_h (y_i | \bm x_i))},
$$
we can write
$$
l_i^{\eta_{ki}} = w_{ki} - \alpha_{ki}.
$$
In order to calculate second derivatives, we need
$$
w_{ki}^{\eta_{ji}}  
= w_{ki}(\delta_{k}^{j}-w_{ji}),
$$
$$
\alpha_{ki}^{\eta_{ji}} 
= \alpha_{ki}(\delta_{k}^{j} - \alpha_{ji}),
$$
where $\delta_k^j = 1$ if $k=j$ and zero otherwise.
Then, we have
$$
l_i^{\eta_{ki} \eta_{ji}}
= w_{ki}(\delta_{k}^{j}-w_{ji})-\alpha{}_{ki}(\delta_{k}^{j}-\alpha{}_{ji})=l^{\eta_{ki}}(\delta_{k}^{j}-w_{ji})-\alpha{}_{ki}l^{\eta_{ji}},
$$
while the third derivatives are
$$
l_i^{\eta_{ki} \eta_{ji} \eta_{mi}}
= (\delta_{k}^{j}-w_{ji})l^{\eta_{ki}\eta_{mi}}-w_{ji}(\delta_{j}^{m}-w_{mi})l^{\eta_{ki}}-\alpha{}_{ki}(\delta_{k}^{m}-\alpha{}_{mi})l^{\eta_{ji}}-\alpha{}_{ki}l^{\eta_{ji}\eta_{mi}}.
$$
We can now write the derivatives of the log-likelihood with respect to the regression coefficients. Given that $l^{\bm \beta^k} = {\mathbf {X}^k}^\top l^{\bm \eta_{k}}$, for $k=2, \dots, K$, the gradient is
\begin{equation}
l^{\bm \beta} = \left( {l^{\bm \beta^2}}^\top, \dots, {l^{\bm \beta^K}}^\top \right)^\top
=
\left( {l^{\bm \eta_{2}}}^\top {\mathbf {X}^2}, \dots, {l^{\bm \eta_{K}}}^\top {\mathbf {X}^K} \right)^\top
.
\end{equation}
Second derivatives are $l^{\bm \beta^k \bm \beta^j} = {\mathbf {X}^k}^\top \mathbf D^{kj} {\mathbf {X}^j}$, with $k,j=2, \dots, K$, where $\mathbf D^{kj}$ is a $N \times N$ diagonal matrix whose diagonal is $l^{\bm \eta_k, \bm \eta_j}$.
Then, we can write the Hessian of the log-likelihood as
\begin{equation}
l^{\bm \beta \bm \beta} = 
\begin{pmatrix}
l^{\bm \beta^2 \bm \beta^2} & \cdots & l^{\bm \beta^2 \bm \beta^K} \\
\vdots & \ddots & \vdots \\
l^{\bm \beta^K \bm \beta^2} & \cdots & l^{\bm \beta^K \bm \beta^K}
\end{pmatrix}
=
\begin{pmatrix}
{\mathbf {X}^2}^\top \mathbf D^{22} {\mathbf {X}^2} & \cdots & {\mathbf {X}^2}^\top \mathbf D^{2K} {\mathbf {X}^K} \\
\vdots & \ddots & \vdots \\
{\mathbf {X}^K}^\top \mathbf D^{K2} {\mathbf {X}^2} & \cdots & {\mathbf {X}^K}^\top \mathbf D^{KK} {\mathbf {X}^K}
\end{pmatrix}
.
\end{equation}

\subsection{Gradient of the LAML w.r.t the log smoothing parameters}

In this section, we show how to compute the gradient of the Laplace approximate marginal likelihood, $\tilde{\mathcal{V}}(\bm \lambda)$, w.r.t the log smoothing parameters, which is required for BFGS optimisation.
In particular, we need to calculate the likelihood specific term $\partial \log \lvert \bm{\mathcal H} \rvert / \partial \rho_g$, where
$\bm{\mathcal H} = - \mathcal L^{\bm \beta \bm \beta} (\hat{\bm \beta})$ is the negative Hessian of the posterior log-density, evaluated at its maximiser, and $\rho_g = \log \lambda_g$.
Note that, as shown in \cite{wood2016smoothing},
$$
\frac{\partial \log \lvert \bm{\mathcal H} \rvert}{\partial \rho_g} = \text{tr} \left( \bm{\mathcal H}^{- 1} \frac{\partial \bm{\mathcal H} }{\partial \rho_g} \right),
$$
$$
\frac{\partial \bm{\mathcal H}}{\partial \rho_g} = - l^{\hat{\bm \beta} \hat{\bm \beta} \rho_g} + \lambda_g \mathbf S_g,
$$
where $l^{\hat{\bm \beta} \hat{\bm \beta} \rho_g} 
= \sum_{i=1}^N l_i^{\hat{\bm \beta} \hat{\bm \beta} \rho_g}$. The element in the $u$-th row and $v$-th column of $l^{\hat{\bm \beta} \hat{\bm \beta} \rho_g}$ is
$$
\left( l^{\hat{\bm \beta} \hat{\bm \beta} \rho_g} \right)_{uv} 
=  l^{\hat{\beta}^{k}_{r} \hat{\beta}^{j}_{s} \rho_g} 
= \sum_{m=2}^K \sum_{t=1}^{p_m}  
l^{\hat{\beta}^{k}_{r} \hat{\beta}^{j}_{s} \hat{\beta}^{m}_{t} } \frac{d \hat{\beta}^m_t}{d \rho_g},
$$
where ${d \hat{\bm \beta}}/{d \rho_g}$ can be obtained by implicit differentation as
$$
    \frac{d \hat{\bm \beta}}{d \rho_g} 
    = \bm{\mathcal H}^{- 1} \lambda_g \textbf S_g \hat{\bm \beta}.
$$
Above, $k$ and $j$ indicate the coefficients vectors to which the $u$-th and $v$-th elements of $\hat{\bm \beta}$ belong (i.e., $\hat{\bm \beta}^k$ and $\hat{\bm \beta}^j$). Similarly, $r$ and $s$ are the indices of the elements of $\hat{\bm \beta}^k$ and $\hat{\bm \beta}^j$ corresponding the $u$-th and $v$-th elements of $\hat{\bm \beta}$. 
In the following we show how to compute $l^{\hat{\bm \beta} \hat{\bm \beta} \rho_g}$ efficiently, that is without explicitly computing all the $p^3$ third order derivatives of $l$ w.r.t. $\hat{\bm \beta}$, where $p = \sum_{k=2}^K p_k$. 

Let us define the following vector of third derivatives of the log-likelihood w.r.t. the linear predictors
$$
l^{\bm \eta_k, \bm \eta_j, \bm \eta_m} = \left( l_1^{\eta_{k1} \eta_{j1} \eta_{m1}}, \dots, l_N^{\eta_{kN} \eta_{jN} \eta_{mN}} \right) ^ \top,  \quad \text{for} \quad k,j,m=2, \dots, K
$$
Third derivatives with respect to the regression coefficients are
$$
l^{\beta_{r}^{k}\beta_{s}^{j}\beta_{t}^{m}} 
= \sum_{i=1}^N l_i^{\eta_{ki}\eta_{ji}\eta_{mi}}X_{{i}r}^{k}X_{{i}s}^{j}X_{{i}t}^{m},
$$
where $X_{{i}r}^{k}$ denotes the element $(i,r)$ of the matrix $\mathbf X^k$.
For each smoothing log-parameter $\rho_g$, with $g=1,\dots, G$, we have
$$
l^{\hat{\beta}_{r}^{k} \hat{\beta}_{s}^{j}\rho_{g}}
=\sum_{m=2}^{K} \sum_{t=1}^{p^m} l^{\hat{\beta}_{r}^{k} \hat{\beta}_{s}^{j} \hat{\beta}_{t}^{m}} \frac{d\hat{\beta}_{t}^{m}}{d\rho_{g}}
=\sum_{i=1}^{N} \sum_{m=2}^{K} \sum_{t=1}^{p^m} l^{\eta_{ki}\eta_{ji}\eta_{mi}} X_{ir}^{k} X_{{i}s}^{j} X_{{i}{t}}^{{m}} \frac{d\hat{\beta}_{{t}}^{{m}}}{d\rho_{g}},
$$
which can be computed in $O(N{p^{(1)}}^2K^2) + O(N p^{(1)} K^3)$, with $p^{(1)}=\max_{h=2,\dots,K}p^h$, by doing, for each $k,j$ block $(k,j=2,\dots,K)$
$$
l^{\hat{\bm \beta}^{k}\hat{\bm \beta}^{j}\rho_{g}} = {\bf X}^{k\top}\mathbf{V}_{g}^{kj}{\bf X}^{j},
$$
where $\mathbf{V}_{g}^{kj}$ is a diagonal $N\times N$ matrix with diagonal elements
$$
(\mathbf{V}_{g}^{kj})_{ii}
= \sum_{m=2}^{K} \sum_{t=1}^{p^m} l^{\eta_{ki}\eta_{ji}\eta_{mi}} X_{i{t}}^{{m}}\frac{d\hat{\beta}_{{t}}^{{m}}}{d\rho_{g}}.
$$
Since $p^{(1)}$ in general is much larger than $K$, the total computational cost of calculating first derivatives of $l^{\hat{\bm \beta} \hat{\bm \beta}}$ with respect to the log smoothing parameters is $O(N{p^{(1)}}^2K^2G)$. Finally, we can write the matrix of derivatives of the Hessian of the log-likelihood with respect to each smoothing parameter $\rho_g$, with $g=1,\dots,G$, as
\begin{equation}
l^{\hat{\bm \beta}\hat{\bm \beta}\rho_{g}} = 
\begin{pmatrix}
l^{\hat{\bm \beta}^{2}\hat{\bm \beta}^{2}\rho_{g}} & \cdots & l^{\hat{\bm \beta}^{2}\hat{\bm \beta}^{K}\rho_{g}} \\
\vdots & \ddots & \vdots \\
l^{\hat{\bm \beta}^{K}\hat{\bm \beta}^{2}\rho_{g}} & \cdots & l^{\hat{\bm \beta}^{K}\hat{\bm \beta}^{K}\rho_{g}}
\end{pmatrix}.
\end{equation}

\section{Approximating the ALE effects' variance via the delta method}
\label{app:ALEdeltaMethod}

For ease of reference, here we define again some variables that have already been defined in the main text. Consider a model, not necessarily a stacking model, with scalar output $\alpha(\bm x)$, where $\bm x$ is a $d$-dimensional vector of model inputs and $\alpha(\bm x)$ is parametrised by the $p$-dimensional vector of model parameters $\bm \beta$. Let ${\bm x}_1, \dots, {\bm x}_N$ be the observed values of $\bm x$ and indicate with ${\bf V}_{\bm \beta}$ the covariance matrix of $\bm \beta$. Let $x_j$ be the $j$-th input variable, which we assume to be continuous and let $x_{i,j}$ be its $i$-th observed value. The case where $x_j$ is a factor variable will be considered at the end of this section. Define a grid $z_{0,j}, \dots, z_{B,j}$ of values along $x_j$, such that $z_{0, j}$ and $z_{B,j}$ are the smallest and the largest observed values of $x_j$. Let $n_j(1), n_j(2), \dots, n_{j}(B)$ be the number of $x_{ij}$'s falling in $[z_{0,j}, z_{1,j}), [z_{1,j}, z_{2,j}), \dots, [z_{B-1,j}, z_{B,j}]$. Indicate with $v_j(x) \in \{1, \dots, B\}$ the bin number in which an arbitrary value $x$ of $x_j$ belongs to and let $S_j(v)$ be the set such that, if ${\bm x}_i \in S_j(v)$, then $x_{i, j}$ belongs to the $v$-th bin. The uncentered ALE effect of $x_j$ is estimated by
$$ 
\hat{\alpha}_{j,ALE}(x)=\sum_{v=1}^{v_{j}(x)}\frac{1}{n_{j}(v)}\sum_{\{i:x_{i,j}\in S_{j}(v)\}}[\alpha(z_{v,j},\bm{x}_{i,\backslash j})-\alpha(z_{v-1,j},\bm{x}_{i,\backslash j})], \;\; \text{with} \;\; \hat{\alpha}_{j,ALE}(z_{0, j}) = 0.
$$
To simplify the notation let us drop the index $j$ from $v_j(x)$, $n_j(v)$, $S_j(v)$, $\bm x_{i,\backslash j}$ and $z_{v,j}$. Define the set of indices $i_{v, 1}, \dots, i_{v, n(v)}$ such that ${\bm x}_{i_{v, 1}}, \dots, {\bm x}_{i_{v, n(v)}} \in S(v)$, for $v = 1, \dots, \tilde{B}$. This allows us to re-express the $j$-th ALE effect as
\[
\hat{\alpha}_{j,ALE}(x)=\sum_{v=1}^{v(x)}\frac{1}{n(v)}\sum_{h=1}^{n(v)}[\alpha(z_{v},\bm{x}_{i_{v,h}})-\alpha(z_{v-1},\bm{x}_{i_{v, h}})].
\]
Define the $n \times d$ matrices
\[
{\bf Z}^1 = \left(\begin{array}{c}
\{z_{0}, \bm{x}_{i_{1, 1}}\}\\
\vdots \\
\{z_{0}, \bm{x}_{i_{1, n(1)}}\} \\
\vdots \\
\{z_{B-1}, \bm{x}_{i_{B, 1}}\} \\
\vdots \\
\{z_{B-1}, \bm{x}_{i_{B, n(B)}}\} \\
\end{array}\right),
{\bf Z}^2 = \left(\begin{array}{c}
\{z_{1}, \bm{x}_{i_{1, 1}}\}\\
\vdots \\
\{z_{1}, \bm{x}_{i_{1, n(1)}}\} \\
\vdots \\
\{z_{B}, \bm{x}_{i_{B, 1}}\} \\
\vdots \\
\{z_{B}, \bm{x}_{i_{B, n(B)}}\} \\
\end{array}\right),
\]
and indicate with $\bm z^1_l$ and $\bm z^2_l$ their $l$-th rows. Define also the $N$-dimensional vectors ${\bf f}^1$ and ${\bf f}^2$ such that $f_l^1 = \alpha({\bm z}^1_l)$ and $f_l^2 = \alpha({\bm z}^2_l)$, the $B \times N$ matrix 
\[
\mathbf{A}=\left(\begin{array}{cccc}
\underbrace{\frac{1}{n_{j}(1)},\dots,\frac{1}{n_{j}(1)}}_{1 \times n_{j}(1)}\\
 & \underbrace{\frac{1}{n_{j}(2)},\dots,\frac{1}{n_{j}(2)}}_{1 \times n_{j}(2)}\\
 &  & \dots\\
 &  &  & \underbrace{\frac{1}{n_{j}(B)},\dots,\frac{1}{n_{j}(B)}}_{1 \times n_{j}(B)}
\end{array}\right),
\]
and indicate with ${\bf 1}_w$ the $B$-dimensional vector such that its first $w$ elements are equal to one and the rest zero. Then the $j$-th ALE effect can be written in matrix form as follows
\[
\hat{\alpha}_{j,ALE}(x)={\bf 1}_{v(x)} \ts {{\bf A}} ({\bf f}^2 - {\bf f}^1).
\]
Let ${\bf J}^1 = \nabla_{\bm \beta} \ts {\bf f}^1$ and ${\bf J}^2 = \nabla_{\bm \beta} \ts {\bf f}^2$ be the $N \times p$ Jacobian matrices of ${\bf f}^1$ and ${\bf f}^2$ w.r.t. $\bm \beta$. Then we have that
\[
\nabla_{\bm \beta} \hat{\alpha}_{j,ALE}(x)={\bf 1}_{v(x)} \ts {{\bf A}} ({\bf J}^2 - {\bf J}^1),
\]
and applying the delta method leads to the approximation
\[
\text{var}\{\hat{\alpha}_{j,ALE}(x)\} \approx \nabla_{\bm \beta}\ts \hat{\alpha}_{j,ALE} \, {\bf V}_{\bm \beta} \, \nabla_{\bm \beta} \hat{\alpha}_{j,ALE}.
\]
Now, let us consider the centered ALE effects, which we estimate by
\[
\hat{\tilde{\alpha}}_{j,ALE}(x)=\hat{\alpha}_{j,ALE}(x)-\frac{1}{N}\sum_{v=1}^{B} n(v) \hat{\alpha}_{j,ALE}\left(\frac{z_{v}+z_{v-1}}{2}\right).
\]
Then we can we use $\hat{\alpha}_{j,ALE}\left\{(z_{v}+z_{v-1})/2\right\} \approx \{\hat{\alpha}_{j,ALE}(z_{v}) + \hat{\alpha}_{j,ALE}(z_{v-1})\}/2$ to derive
\[
\nabla_{\bm \beta} \ts \hat{\tilde{\alpha}}_{j,ALE}(x) \approx \nabla_{\bm \beta} \ts \hat{\alpha}_{j,ALE}(x) - \frac{1}{2N}\sum_{v=1}^{B} n(v) \left\{\nabla_{\bm \beta} \ts \hat{\alpha}_{j,ALE}(z_v) + \nabla_{\bm \beta} \ts \hat{\alpha}_{j,ALE}(z_{v-1})\right\},
\]
which is the main ingredient needed to estimate $\text{var}\{\hat{\tilde{\alpha}}_{j,ALE}(x)\}$ via the delta method.

So far we have assumed that $x_j$ is a continuous variable. If $x_j$ is a factor variable, then 
$z_0, \dots, z_{B}$ represent the unique $B+1$ values of $x_j$, ordered as suggested in Appendix E of \cite{apley2016visualizing}, while $n(v)$ with $v=0, \dots, B$ represents the number of $x_{ij}$'s that have taken the value $z_v$. Then, the uncentered ALE effects are defined by
\begin{align*}
\hat{\alpha}_{j,ALE}(x) = \sum_{v=1}^{v(x)}\frac{1}{n(v)+n(v-1)}& \Big\{ \sum_{h=1}^{n(v)}[\alpha(z_{v},\bm{x}_{i_{v,h}})-\alpha(z_{v-1},\bm{x}_{i_{v, h}})] + \\ &
\sum_{h=1}^{n(v-1)}[\alpha(z_{v},\bm{x}_{i_{v-1,h}})-\alpha(z_{v-1},\bm{x}_{i_{v-1, h}})]\Big\}.
\end{align*}
with $\hat{\alpha}_{j,ALE}(z_0) = 0$. The extra term $\alpha(z_{v},\bm{x}_{i_{v-1,h}})-\alpha(z_{v-1},\bm{x}_{i_{v-1, h}})$ is there because the observations fall only on the $z_v$'s, not between them as in the continuous case, hence we average the differences between the effects at $z_v$ and $z_{v-1}$ by fixing the other variables both at $v$ and at $v-1$. The centred ALE effects are defined similarly to the continuous case, with $z_v$ in place of $(z_v + z_{v-1})/2$. The gradient of either centred or uncentred ALE factor effects are derived similarly to the continuous case. 

As shown in this section, applying the delta method to approximate the variance of the ALE main effects requires the Jacobian of the model output w.r.t. the parameters $\bm \beta$. In the next section we provide the Jacobian under the multinomial parametrisation used in additive stacking.

\subsection{Jacobian under the multinomial parametrisation}

Let  ${\alpha}_{k}$, for $k=1, \dots, K$, be the weights attached to the experts in additive stacking. These are linked to the linear predictors, ${\eta}_{1}, \dots, {\eta}_K$, via the multinomial parametrisation, that is
\[
{\alpha}_{k}=\frac{\exp{\eta}_{k}}{\sum_{a=1}^{K}\exp{\eta}_{a}},
\]
where $\eta_1 = 0$ for identifiability. Let $\bm{\eta}_{k} = \{\eta_{k1}, \dots, \eta_{kN}\}$ be the vector containing the values of the $k$-th linear predictor at each observation. For $k=2, \dots, K$, we have that $\bm{\eta}_{k}=\mathbf{X}^{k}\bm{\beta}^{k}$ where ${\bf X}^k$ and $\bm \beta^k$ are, respectively, the $N \times p_k$ model matrix and the $p_k$-dimensional vector of regression coefficients belonging to the $k$-th linear predictor. Impose $\bm \eta_1 = {\bm 0}$ with $p_1 = 0$ and define ${\bm \alpha}_{k} = \{{\alpha}_{k}(\bm x_1), \dots, {\alpha}_{k}(\bm x_N)\}$. Then, the $N \times p$ Jacobian matrix of $\bm \alpha_k$ w.r.t. $\bm \beta$ is
\[ 
\mathbf{J}= \nabla_{\bm \beta}\ts \bm \alpha_k = \{ \nabla_{\bm \beta^2}\ts \bm \alpha_k, \cdots, \nabla_{\bm \beta^{K}}\ts \bm \alpha_k\},
\]
where $p = \sum_{k=1}^{K} p_k$. By the chain rule, we have 
$$
\nabla_{\bm \beta^a}\ts \bm \alpha_k = \nabla_{\bm\eta_{a}} \ts \bm \alpha_{k} \nabla\ts_{\bm \beta_a} {\bm \eta_a} = \nabla_{\bm\eta_{a}} \ts \bm \alpha_{k} {\bf X}^a,
$$
where $\nabla_{\bm\eta_{a}} \ts \bm \alpha_{k}$ is an $N \times N$ diagonal matrix with non-zero entries
\[
\left(\nabla_{\bm\eta_{a}} \ts \bm \alpha_{k} \right)_{ii} = \frac{\partial\alpha_{ki}}{\partial\eta_{ai}}=\alpha_{ki}\{\mathbbm{1}(k = a)-\alpha_{ai}\},
\]
for $a = 2, \dots, K$ and $k = 1, \dots, K$.

\section{Details on the household demand forecasting application} \label{app:detailsOnApplication}

\subsection{Data preparation}

Data have been filtered in order to exclude customers for which the demand data were not considered interesting for the proposed application. 
In particular, we did not consider customers for which the 99th quantile of the electricity demand over the entire year is less than 0.4kWh. 
Moreover, since the demand of some customers was constant along most of the year and was not considered interesting for the forecasting application, we excluded from the analysis also all customers for which the vector of differences of consecutive demand values contained more than 2500 zeros, over the entire year.
We ended up with a data set of $2565$ customers.
We also removed from the days corresponding to national holidays because, in an operational setting, forecasting electricity demand during these periods requires manual intervention, as demand behavior is anomalous relative to the rest of the year. 
In particular, we excluded days of the year equal to 1, 2 (first two days of the year), 87 (Sunday before Easter), 94, 95, 96 (Easter and two following days), 120, 121, 122 (May Day and two days before), 143, 144, 145 (Pentecost Monday), 304 (Halloween), 358, 359, 360 (24, 25, and 26 December), 365 (New Year's Eve).

The temperature data from NCEI was measured at ten different locations in Ireland. We built a single temperature variable by averaging these temperatures with uniform weights. Note that, when we forecast load one day ahead under the $GaulssInd$ and $GaulssCommon$ model, we use the observed temperatures over that day to compute the smoothed temperatures $T^s_i$. In an operational setting future temperatures would not be available, and a
forecast would be used instead. However, considering that the experts use smoothed temperatures, which strongly depend on the past, and that temperature typically has a much weaker effect on household demand than on aggregate data, we expect that substituting the observed temperatures with a forecast would have a very minor effect on the results presented in the main paper.

\subsection{Experts and additive stacking}

In this section, we provide additional details about the experts used in the additive stacking. First note that, when predicting observation $y^c_i$ for customer $c$, lag values $y^c_{i - 48k}$ are not always available because we excluded some days in the data set. For this reason, when we have missing data $y^c_{i - 48k}$ at a given day of the year $j_{i - 48k}$, with substitute it with the most recent observation available before day $j_{i - 48k}$, at the same time of the day.
When calculating the smoothed temperature $T_i^s = \alpha T_{i-1}^s + (1 - \alpha) T_i$, if $T_{i-1}$ is missing we set $T_i^s = T_i$.  We assume that, for physical reasons, the electricity demand $y_i$ cannot exceed 20 kWh, hence we truncate and re-normalise all probabilistic forecasts from each expert accordingly. 

We provide details about smooth effects used in each expert below.
In the $GaulssInd$ expert, $f_1$, $f_2$ and $f_4$ are smooth effect constructed using cubic regression splines basis, where the dimension of the basis is 10 and we penalise the integrated squared first derivative to avoid problems when extrapolating outside the range of the observed covariates in the training data set.
$f_3$ and $f_5$ are cyclic cubic regression splines, where we chose the dimensions of the basis equal to 30 and 20, respectively.
In $Dynamic$, the smooth effect of the time of the day is represented as cyclic cubic regression splines with the dimensions of the basis equal to 10.
In $GaulssCommon$, $f_1$ and $f_4$ are represented as cyclic cubic regression splines with the dimensions of the basis equal to 20, while $f_2$ and $f_3$ are represented as penalised cubic regression splines, where the dimension of the basis are 20 and 10, respectively. Moreover, for $f_2$ we penalise the integrated squared first derivative to avoid problems when extrapolating outside the range of the observed covariates in the training data set.

Regarding the smooth effects used in the additive stacking, in Equation \eqref{eq:stackModelIrish} $f_1$, $f_2$, $f_3$ and $f_4$ are represented as penalised cubic regression splines with the dimensions of the basis equal to 20 for $f_1$ and $f_4$, and equal to 5 for $f_2$ and $f_3$.      

\subsection{Visualisation of covariates effects using accumulated local effect plots}

In this section, we provide ALE plots for all possible expert/covariate pairs based on the final stacking model fitted using data from weeks 9 to 50.
The code for producing ALE plots for additive stacking models is available through the \verb|mgcViz| R package \citep{fasiolo2018scalable}.
Since we are dealing with a big data set, calculation of accumulated local effects is computationally expensive. 
Hence, we obtain them by sub-sampling $10^6$ observations from the training data set (we have checked that the plots do not vary between sub-samples).

\begin{sidewaysfigure}[ht]
    \includegraphics[height=0.5\textheight]{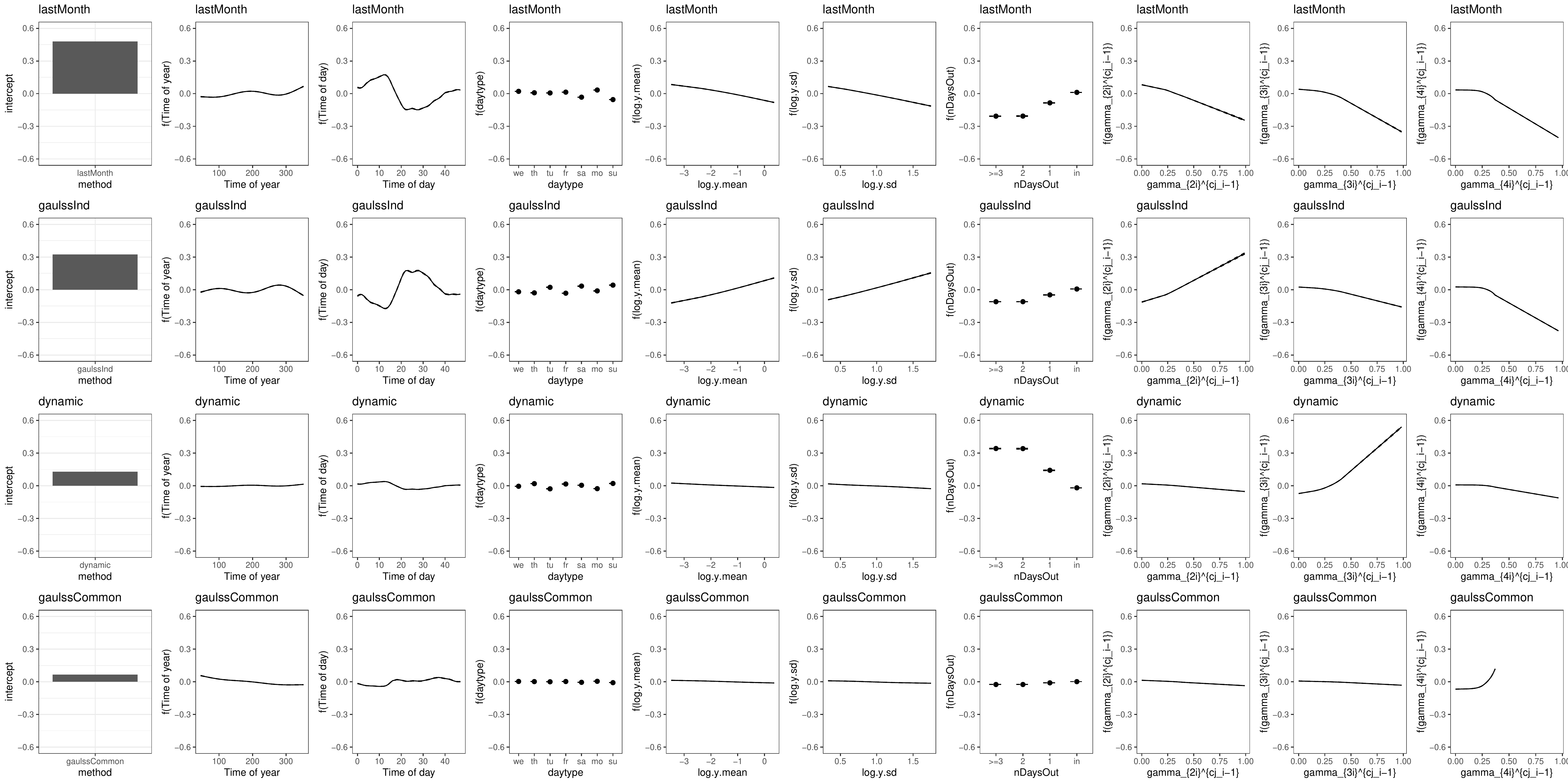}
    \caption{Accumulated local effect plots of covariates on stacking weights (1/2).
    Note that the vertical scale has been changed only for the effect of the covariate $\gamma_{4i}^{cj_i-1}$ on the weight of the expert `GaulssCommon', for which limits have been changed from $(-0.5,0.5)$ to $(-1,1)$ for visualisation purpose.}
    \label{fig:ale1}
\end{sidewaysfigure}

\begin{sidewaysfigure}[ht]
    \includegraphics[height=0.5\textheight]{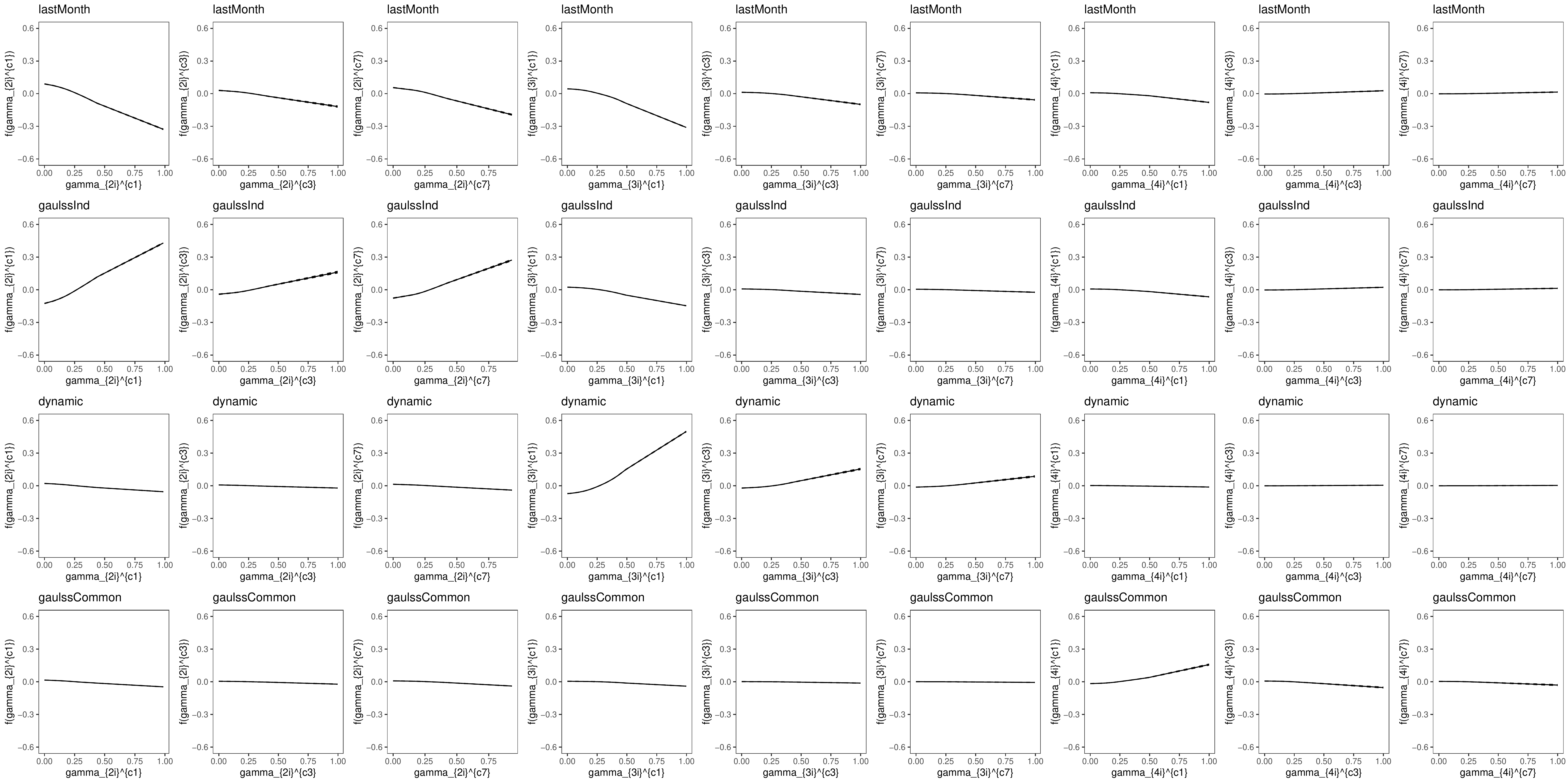}
    \caption{Accumulated local effect plots of covariates on stacking weights (2/2).}
    \label{fig:ale2}
\end{sidewaysfigure}

Figures \ref{fig:ale1} and \ref{fig:ale2} show the centered ALE effects of the covariates on stacking weights.
Each row in the two figures shows plots related to one expert, while each column refers to a single covariate. 
The intercept shows that, on average, the \textit{LastMonth} expert has the highest weight in the mixture, while \textit{GaulssCommon} has the smallest weight.
However, the latter is still useful in cases where all other experts fail due to some important change in the electricity demand. For example, a customer may have near-zero consumption at the beginning of the year and then a sudden increase in demand. Such a step change may dramatically affect the predictive performance of the customer-specific experts, which have never seen high demand values (because they are fitted separately for each customer).
In those cases, the \textit{GaulssCommon} expert, which borrows information across all the customers, often produces more robust predictions.

The ALE plots in Figure \ref{fig:ale1} show that, in general, the most complex expert, \textit{GaulssInd}, is given more weight in the mixture when demand dynamics are complex. 
In particular, it obtains more weight for customers with higher average electricity demand, as well as higher standard deviation.
The effects of the time of the day and day of the week also confirm the higher importance of \textit{GaulssInd} in the mixture when demand volatility is higher. In particular, the expert has a larger weight during daytime, when demand dynamics are more complex, and during weekends, when the demand is less predictable.

As expected, the weight of the \textit{Dynamic} expert depends strongly on the $\text{do}_i^c$ variable, denoted as nDaysOut in the figure, which counts for how  many days the customer has been out of home before the current day.
When a customer leaves home for a long period, e.g. for holidays, all other experts perform badly because they rely heavily on historical data, while \textit{Dynamic} does better because it reacts quickly on the basis of the most recent data.

The last three columns in Figure \ref{fig:ale1} show the effects of the covariates $\gamma_{ki}^{cj_i-1}$, $k=2,3,4$, and show that experts that perform better than the other experts for a customer on all the historical data available up to that point tend to have a higher weight in the mixture. This particular effect is strongest for the  \textit{GaulssCommon} experts (see the effect of $\gamma_{4i}^{cj_i-1}$ on its weight). The remaining plots in Figure \ref{fig:ale2} show the accumulated local effect plots for covariates $\gamma_{ki}^{cd}$, $k=2,3,4$, $d=1,3,7$, evaluating the relative log-score performance of the expert $M_k$ with respect to the others in the last $d$ days, as explained in Section \ref{sec:stacking_model}.
As expected, the weight of the $k$-th expert increases with the corresponding score, $\gamma_{ki}^{cd}$. An exception is \textit{GaulssCommon}, whose weight slightly decreases with $\gamma_{4i}^{c3}$ and $\gamma_{4i}^{c7}$.

\end{document}